# Dependence of energy relaxation and vibrational coherence on the location of light-harvesting chromoproteins in photosynthetic antenna protein complexes

**Masaaki Tsubouchi,[1,2,a] Nobuhisa Ishii,[1] Takatoshi Fujita,[2] Motoyasu Adachi,[2] and Ryuji Itakura[1]**

## AFFILIATIONS

[1]Kansai Institute for Photon Science (KPSI), National Institutes for Quantum Science and Technology (QST), 8-1-7 Umemidai, Kizugawa, Kyoto 619-0215, Japan

[2]Institute for Quantum Life Science, National Institutes for Quantum Science and Technology (QST), 4-9-1 Anagawa, Inage, Chiba 263-8555, Japan

[a]**Author to whom correspondence should be addressed**:
tsubouchi.masaaki@qst.go.jp

## ABSTRACT

Phycobilisomes are antenna protein complexes in cyanobacteria and red algae. In phycobilisomes, energy transfer is unidirectional with an extremely high quantum efficiency close to unity. We investigate intraprotein energy relaxation and quantum coherence of constituent chromoproteins of allophycocyanin (APC) and two kinds of C-phycocyanin (CPC) in phycobilisomes using two-dimensional electronic spectroscopy (2D-ES). These chromoproteins have similar adjacent pairs of pigments α84 and β84, which are excited to delocalized exciton states. However, the kinetics and coherence of exciton states are significantly different from each other. Even CPCs with almost the same molecular structure display significantly different spectra and kinetics when the locations in the phycobilisome are different. This difference may be one of the key mechanisms for the efficient and unidirectional energy transfer in phycobilisomes. We observe low-frequency coherent vibrational motion of approximately 200 cm$^{-1}$ with large amplitude and a decay time of 200 fs. The wave packet motion involving energy relaxation and oscillatory motions on the potential energy surface of the exciton state is clearly visualized using beat-frequency-resolved 2D-ES.

## I. INTRODUCTION

Energy transfer in light-harvesting antenna protein complexes is unidirectional with an extremely high quantum efficiency close to unity.[1] This has been investigated via structural and spectroscopic analysis for more than thirty years to understand its mechanism. The interaction between the nearby pigments contained in proteins that generates the delocalized electronic excited states (the exciton states) and the quantum coherence generated in the pair of exciton states have both been discussed. Whether coherence is one of the keys for unidirectional efficient energy transfer has been a topic for concern over the last two decades.[2-14] But despite long-term efforts, the contribution of coherence to energy transfer has not yet been consistently well explained.

Phycobilisomes are the antenna protein complexes in cyanobacteria and red algae living in various water environments.[15] In recent years, the energy transfer in phycobilisomes has been intensively investigated via structural insights obtained by cryo-electron microscopy (cryo-EM).[16-19] Ultrafast non-linear spectroscopies have also revealed energy transfer and quantum coherence based on the knowledge of the structures of phycobilisomes.[20-23] Figure 1(a) shows the cryo-EM structure of *Anabaena* sp. PCC 7120 phycobilisome,[16] which has a structure similar to the *Thermosynechococcus elongatus* BP-1 phycobilisome that is used later in this study. The phycobilisome complex consists of multiple chromoproteins: allophycocyanin (APC) and C-phycocyanin (CPC). The core of the complex is composed of lateral APC assemblies and is surrounded by CPC rods. Both chromoproteins are connected by a linker protein. The light energy received by the phycobilisome is transferred to the photosystem via the core. In the energy transfer pathway, CPC and APC function as the energy donor and acceptor, respectively.

Figures 1(b) and (c) show the structure and the pigment arrangement of the APC trimer and the CPC hexamer, respectively, which are used in this study. The APC monomer consists of the α and β subunits, which contain the phycocyanobilin pigments α84 and β84, respectively. When an APC trimer is formed, the α84 and β84 in neighboring monomers are close together, with a distance of 2.1 nm.[24] The CPC monomer contains the β153 pigment in addition to α84 and β84, and a CPC hexamer is formed. A pair of neighboring α84 and β84 pigments are coupled to form exciton states from electronically excited states of each pigment, as shown in Fig. 1(d). The existence of the exciton states has been confirmed by circular-dichroism spectroscopy.[25, 26] The β153 pigment is decoupled from the pair of α84 and β84, and the energy level of its electronic excited state is higher than those of the exciton states (Fig. 1(e)). The delocalization of the electronic wave function between α84 and β84 due to exciton coupling is considered to contribute to the unidirectional property of the energy transfer, and it has been investigated in both experimental and theoretical studies.[10, 11, 25, 27-35]

The site energy difference and the dipole–dipole interaction between the α84 and β84 pigments of APC are estimated to be $\Delta E$ = 760 $cm^{-1}$ and $J$ = 150 $cm^{-1}$, respectively.[29, 34] The energy splitting between the pair of exciton states is clearly seen in the absorption spectrum, as shown in Fig. 1(f). The peak and the shoulder are observed at 15400 $cm^{-1}$ and 16200 $cm^{-1}$, which correspond to the lower ($e^+$) and higher ($e^-$) exciton states, respectively. The sub-picosecond ultrafast dynamics of APC have been explored by measuring and calculating energy relaxation and quantum coherence.[10, 27-29] The fastest step is the interexciton-state relaxation with the vibrational energy redistribution at a time constant of 30 fs during dephasing of the electronic and nuclear coherence. The change in geometric configurations of the pair of pigments occurs within a time constant of 200–300 fs after the first step mentioned above, followed by the localization of energy. This process has been considered as a key to unidirectional energy transfer. The vibrational mode promoting the energy relaxation is suggested to be the hydrogen out-of-plane wagging motion with a frequency of 800 $cm^{-1}$.[10] This frequency is close to the site energy difference between pigments; therefore, it is considered that the vibronic states with this promoting mode contribute to the delocalized exciton states. To confirm the contribution of the exciton states, the ultrafast dynamics in an α-subunit of the APC monomer that does not contain any pair of pigments were measured and compared to that of the APC trimer.[25, 30] Significant differences in energy relaxation and coherent phenomena were found between the α-subunit and the trimer and, therefore, it is suggested that the exciton interaction between the pigments plays a crucial role in efficient energy relaxation.

The structure of CPC is almost identical to that of APC except for the existence of the β153 pigment. However, the absorption spectrum and the ultrafast dynamics of CPC are significantly different from those of APC.[32-34] As shown in Fig. 1(f), a shoulder-like structure is not clearly observed in the absorption spectrum, and only the peak is seen at around 16100 $cm^{-1}$. This is due to the overlap of the absorption spectra of β153 and the exciton states with a small site energy difference of 350 $cm^{-1}$ between α84 and β84 in CPC, which is less than half the energy difference in APC.[32] The time constant of the energy relaxation to the delocalized state is 970 fs in CPC, which is four times slower than in APC.

In this study, we investigate energy relaxation and quantum coherence in the APC trimer and two kinds of CPC hexamer in *Thermosynechococcus elongatus* BP-1 phycobilisome. The locations of these three chromoproteins are shown in Fig. 1(a). In this study, the CPC hexamer located at the outermost part of the antenna protein complex is called CPC1, while that located between CPC1 and the APC trimer is called CPC2. The polypeptides for structural formation of phycobilisome are different between CPC1 and CPC2. As shown in Fig. 1(f), the peak in the absorption spectrum of CPC2 is slightly red-shifted from that of CPC1, suggesting that CPC2 may receive energy from CPC1. Two-dimensional electronic spectroscopy (2D-ES) is employed to simultaneously measure energy relaxation and quantum coherence from various initial excited states of chromoproteins. The 2D-ES can map the correlation between the initial excited states at the moment of photo-excitation and the states populated at a delay time after photo-excitation.[5, 7, 13, 36-44] The intramolecular energy relaxation and quantum coherence in the three constituent chromoproteins are investigated to discuss the roles of these proteins in efficient energy transfer in phycobilisomes.

## II. EXPERIMENTAL PROCEDURE

In the following, unless otherwise specified, APC and CPC refer to the APC trimer and the CPC hexamer, respectively. The light-harvesting proteins APC, CPC1, and CPC2 were designed on the basis of the structure determined by cryo-EM.[19] The sample preparation and characterization methods are explained in Supplementary Secs. 1 and 2, respectively. Although the synthesized APC was designed to be a hexamer, an APC trimer was prepared instead probably because of an unbalanced expression or unstable hexamer formation. The proteins were diluted with a buffer solution to prepare a sample solution including 10% (w/v) glycerol with protein concentrations ranging from 4.3 to 5.0 mg/mL. The sample solution was circulated by a peristaltic pump through a sample cell with 1 mm thick quartz windows and an optical path length of 0.2 mm. The optical density of the sample solution was OD ≈ 0.4 at the peak of the absorption spectrum, as shown in Fig. 1(f).

The two-dimensional (2D) electronic spectra were measured using a previously reported setup,[45] except that the fully non-collinear BOXCARS geometry was replaced with a pump–probe configuration. An advantage of the latter configuration is that the absorptive 2D electronic spectra directly visualizing the transient behaviors can be obtained automatically with extremely high phase stability.[46] Experimental details are presented in Supplementary Sec. 3 and Fig. S5. In brief, we produced sub-10-femtosecond visible pulses using a Yb:KGW laser followed by two pulse compression stages.[45] The coherent time interval, $\tau$, between two pump pulses was scanned by the Translating-Wedge-Based Identical Pulses eNcoding System (TWINS)[47, 48] from −60 to 60 fs in steps of 0.4 fs. The waiting time, $T$, was scanned from 0 to 2000 fs in steps of 4 fs. The energy density of the excitation pulse irradiating the sample was 40 μJ/cm$^2$ in each pulse, which did not induce a non-linear response by a single pulse, as shown in Supplementary Fig. S6. At the sample position, the linear polarizations of the two laser pulses were set in parallel. The duration of the excitation pulses was 13 fs at the sample position. The spectral width was sufficient to excite the entire absorption spectrum of CPC and APC, as shown in Fig. 1(f). One measurement, which collects ~500 2D electronic spectra, was completed within three hours, and so multiple measurements with several proteins could be performed on the same day, which allowed quantitative comparison between proteins.

## III. RESULTS AND DISCUSSION

### A. Time-resolved 2D electronic spectra

Figure 2(a) shows a series of 2D electronic spectra $S(E_{\mathrm{exc}}, E_{\mathrm{det}}\!: T)$ of APC, which is the acceptor protein involved in energy transfer in phycobilisomes. The horizontal and vertical axes show the excitation ($E_{\mathrm{exc}}$) and detection ($E_{\mathrm{det}}$) energies, respectively. The positive amplitude (red in the 2D electronic spectra) represents the grand state bleaching (GSB) or the stimulated emission (SE) of the excited state. On the other hand, the negative amplitude (blue in the 2D electronic spectra) represents the excited-state absorption (ESA). At the moment of the excitation ($T$ = 0 fs), the signal can be seen in the diagonal line. This indicates that APC is detected at the same energy as the excitation energy before energy relaxation. The lower (15400 cm$^{-1}$) and higher (16200 cm$^{-1}$) exciton states are resolved in the diagonal line. As the pump–probe delay time increases, the higher exciton state relaxes to the lower state. Within a delay time of $T$ = 1 ps, the amplitude distribution in the 2D spectrum is aligned parallel to the horizontal axis, which means that both the lower and higher exciton states almost completely relax to the bottom of the potential energy surface of the lower exciton state. The detection energy in the signal spectra converges to $E_{\mathrm{det}}$ = 15150 cm$^{-1}$, which corresponds to the peak of the fluorescence spectrum of APC with a Stokes shift of 250 cm$^{-1}$.[31]

A series of 2D electronic spectra of the CPC1 energy-donor protein is shown in Fig. 2(b), while that of CPC2 is shown in Supplementary Fig. S7. In both CPC proteins, the positive signals spanning a broad detection-energy range higher than 15300 cm$^{-1}$ are detected at $T$ = 1 ps, unlike in the case of APC. This result indicates that relaxation to the bottom of the potential is not completed and, therefore, the excited vibronic states still survive at picosecond time delays after the excitation. In CPC1, the ESA component also appears at a detection energy below $E_{\mathrm{det}} < 15300\ \mathrm{cm}^{-1}$, which is not clearly seen in the case of APC. As seen in Figs. 1(b) and (c), both APC and CPC1 have similar structures, except for the β153 pigment existing in CPC1. Nevertheless, the absorption spectrum and the energy relaxation kinetics of CPC1 are found to be far different from those of APC. This difference might be related to the efficient energy transfer in phycobilisomes.[33]

Figure 3 summarizes the transient spectra at the long delay time ($T$ = 2 ps) and the absorption spectra. The transient spectra are obtained by projection of the 2D electronic spectra onto the detection energy axis. Because the CPC1 is located at the periphery of the phycobilisome, we assume that the energy transfer in a phycobilisome is initiated by light absorption by CPC1 whose absorption spectrum has a peak at 16100 $\mathrm{cm}^{-1}$ and a broad bandwidth wider than 2000 $\mathrm{cm}^{-1}$. The initially excited vibronic states relax to the lower exciton state. The population of vibronic states after relaxation is shown as the thin red dashed line in Fig. 3, which has a peak at 15600 $\mathrm{cm}^{-1}$. Because of energy relaxation by 500 $\mathrm{cm}^{-1}$, the transient spectra of CPC1 well overlaps with the absorption spectrum of APC with a peak at 15400 $\mathrm{cm}^{-1}$ (black solid line in Fig. 3). This is a well-known condition for the Fluorescence resonance energy transfer (FRET) mechanism.[1] In previous studies, it was reported that the energy transfer between CPCs[20, 21] and between CPC and APC[21] is very fast, with time constants of several hundred femtoseconds to few picoseconds. The APC can be excited to both the higher and lower exciton states by receiving energy from the CPCs in this short time scale. The vibronic states in APC then quickly relax to the bottom of the potential energy surface of the lower exciton state as shown in the transient spectra indicated by the thin black dashed line, which has a peak at 15130 $\mathrm{cm}^{-1}$. Note that the transient spectra of CPC2 which are located between CPC1 and APC, are slightly red shifted from that of CPC1. This may promote the efficient energy transfer in the phycobilisome.

## B. Dynamic Stokes shift and wave packet motion

To visualize the relaxation of energy from a specific initial excited state and the quantum coherence related to that state, a series of 2D electronic spectra $S(E_{\mathrm{exc}}, E_{\mathrm{det}}\colon T)$ is reconfigured into a series of time-resolved detection signal spectra $S(E_{\mathrm{det}}, T\colon E_{\mathrm{exc}})$ at a specific excitation energy, which are similar to the transient absorption (TA) spectra. In conventional TA spectroscopy, time resolution is sacrificed to improve the energy resolution of the pump light.

In contrast, in 2D-ES, time resolution is ensured by measuring the correlation between the pump and probe energies. Figures 4(a) and (b) plot $S(E_{\mathrm{det}}, T{:}\, E_{\mathrm{exc}})$ of APC and CPC1, respectively. Those of CPC2 are shown in Supplementary Fig. S8(b). No matter what energy the APC was excited with, it relaxed to the lowest energy level ($E_{\mathrm{det}}$ = 15150 cm$^{-1}$) within a few hundred femtoseconds. On the other hand, in CPC1, the high-energy component remains at the long delay time after excitation.

Figure 4(c) is an enlarged view of $S(E_{\mathrm{det}}, T{:}\, E_{\mathrm{exc}})$ of APC at $E_{\mathrm{exc}}$ = 15500 cm$^{-1}$, which is the peak energy in the absorption spectrum. In addition to the monotonous energy relaxation, the oscillatory component can be clearly found. To analyze the oscillatory component quantitatively, the detection energies with the maximum amplitude at each pump–probe delay, $T$, are plotted in Fig. 4(c) as a white solid line. The dynamic Stokes shift from 15500 cm$^{-1}$ to 15150 cm$^{-1}$ is rapidly completed with a time constant of 62 fs. The beat signal is clearly seen in the residues of the fitting by the single exponential decay function as shown in Fig. 4(e). The amplitude of the modulation in the first oscillation is ±95 cm$^{-1}$, which is 27% of the dynamic Stokes shift. The decay time of the oscillatory component is 192 fs, which is three times longer than the time constant of the dynamic Stokes shift. The derived beat frequency is $\nu_{\mathrm{beat}} =$ 213 cm$^{-1}$, as shown in the Fourier transform spectrum of the residue time profile (Fig. 4(f)). This beat signal indicates that the vibrational wave packet is generated on the excited electronic state by the pump laser light.

Figure 4(d) shows an enlarged view of $S(E_{\mathrm{det}}, T{:}\, E_{\mathrm{exc}})$ of CPC1 at $E_{\mathrm{exc}}$ = 16200 cm$^{-1}$, which is the peak energy in the absorption spectrum (Fig. 1(f)). The rapid dynamic Stokes shift and the amplitude modulation with respect to the pump–probe delay time can also be seen in the map of CPC1. However, the oscillation of the wave packet cannot be visualized in the time profile of the detection energy with the maximum amplitude, unlike for the APC (see the solid white line in Fig. 4(d)). Details of the difference in the oscillatory behavior between APC and CPC are discussed in Sec. III.D with the beat-frequency-resolved 2D electronic spectra.

## C. Energy relaxation and quantum coherence

Figure 5(a) shows the signal amplitudes of APC as a function of the pump–probe delay time at specific coordinates $(E_{\mathrm{exc}}, E_{\mathrm{det}})$ in the 2D electronic spectra as shown in Fig. 2. The excitation energy is 15500 cm$^{-1}$ which is close to the peak energy of the absorption spectrum. Around zero-time delay, $T \approx 0$ fs, extremely rapid changes are seen in some of the time profiles. These are due to mixing of contributions from ultrafast kinetics, quantum beat, and the coherent spike. Since it is difficult to separate these three components, in the following we will avoid discussing phenomena that occur within a few tens of femtoseconds after excitation. At detection energies higher than $E_{det}$ = 15250 cm$^{-1}$, the single exponential decay profiles superimposed with the quantum beat are observed, while at the lower detection energies, the exponential rise profiles with quantum beat are observed. The quantum beat at detection energies above the boundary (around $E_{det}$ = 15200 cm$^{-1}$) is out of phase with that at the energies below the boundary. Note that the quantum beat is barely seen in the time profiles around the boundary energy. This is a typical feature of a vibrational wave packet that oscillates on the potential energy surface of an electronically excited state while relaxing its energy to the potential bottom. It will be explained in detail later using a beat-frequency-resolved 2D electronic spectrum in Sec. III.D.

The time profiles of the signal amplitude for CPC1 are shown in Fig. 5(b). The excitation energy is close to the peak energy of the absorption spectrum, 16150 cm$^{-1}$. At all detection energies, from $E_{det}$ = 15300 to 16200 cm$^{-1}$, a slow decay component with a time constant of larger than 1 ps is observed. The quantum beat can be observed in the profiles, but the amplitude and the decay time of the beat signals are much smaller and shorter than those in APC, respectively. This implies that decoherence due to interaction with the solvent(s) and/or the other pigments, more rapidly occurs in CPC1 than in APC.

When the excitation to the higher exciton state with an excitation energy of $E_{exc}$ = 16650 cm$^{-1}$ occurs, the quantum beat is barely observed in the time profiles, as shown in Figs. 5(c) and (d) for APC and CPC1, respectively. In the case of APC, the boundary between the rising and decaying profiles can be seen around $E_{det}$ = 15250 cm$^{-1}$, which is similar to that for the case of excitation to the lower exciton state of APC with an energy of $E_{exc}$ = 15500 cm$^{-1}$. All signals come from either GSB or SE because the signal amplitudes are all positive. On the other hand, the time profile of CPC1 involves only the population decay with the positive ($E_{det}$ > 15500 cm$^{-1}$) or negative amplitude ($E_{det}$ < 15250 cm$^{-1}$). The positive and negative signals correspond to the detection processes of GSB (or SE) and ESA, respectively.

### D. Beat-frequency-resolved 2D electronic spectra

For a quantitative evaluation of energy relaxation and quantum coherence, the quantum beat is extracted from the observed time profiles by global fitting as described in Supplementary section 7. First, the time profiles at all the grid points of $(E_{\mathrm{exc}}, E_{\mathrm{det}})$ in the series of the time-resolved 2D electronic spectra $S(E_{\mathrm{exc}}, E_{\mathrm{det}}, T)$ are fitted to the double-exponential functions as,

$$S(E_{\rm exc}, E_{\rm det}, T) = A_{\rm f}(E_{\rm exc}, E_{\rm det}) \exp\{-T/\tau_{\rm f}(E_{\rm exc})\}$$
$$+A_{\rm s}(E_{\rm exc}, E_{\rm det}) \exp\{-T/\tau_{\rm s}(E_{\rm exc})\} + A_{\rm c}(E_{\rm exc}, E_{\rm det}), \quad (1)$$

where $\tau_{\rm f}(E_{\rm exc})$ and $\tau_{\rm s}(E_{\rm exc})$ are the time constants obtained by global fitting to all the time profiles with the same excitation energy $E_{\rm exc}$. The results of the fitting are shown in Fig. 5 as the black solid lines. Because the ultrafast response within a few tens of femtoseconds is hardly separated from the oscillatory behavior of the quantum beat and the coherent spike, it is difficult to precisely determine the time constant of the fastest step $\tau_{\rm f}(E_{\rm exc})$ regarding to the internal conversion between the interexciton-states as reported before.[10, 27-29] Therefore, we analyze the time profiles after a delay time of 50 fs in the following discussions.

Figures 6(a) and (b) show the beat-frequency spectra for APC and CPC1, respectively. in intensity. The excitation energies are the peak of the absorption spectra at 15500 cm$^{-1}$ for APC and 16150 cm$^{-1}$ for CPC1. These spectra are calculated by Fourier-transforming the time profiles of the quantum beat shown in Supplementary Fig S12 which are obtained as the residuals of the fitting. The strong peaks are found at around 200 cm$^{-1}$. This mode was identified to be isotropic motion according to anisotropy measurements through transient grating spectroscopy[10, 31] and 2D-ES.[25] However, this low-frequency mode has not been assigned to a specific purely vibrational or vibronic mode. The two sharp peaks are resolved at $\nu_{\rm beat} = 202$ and 266 cm$^{-1}$ for CPC1 in Figs. 6(b). In contrast, only a single broad peak can be seen around 200 cm$^{-1}$ in the intensity spectra of APC in Fig. 6(a). On the other hand, in the real part of the complex spectra of APC shown in Fig. 6(c), the two peaks are resolved at $\nu_{\rm beat} = 176$ and 214 cm$^{-1}$ with opposite amplitude. The amplitude of the peak at 214 cm$^{-1}$ flips from negative to positive as the detection energy decreases from 15700 to 14800 cm$^{-1}$.

In the time-dependent 2D electronic spectra (Fig. 2), the quantum beat signal is usually overlapped with the incoherent population background. The beat-frequency-resolved analysis of 2D electronic spectra has been proposed to illustrate the coherence amplitude maps between the quantum states.[45, 49-51] Figures 7(a) and (b) show 2D coherent maps of APC in intensity at beat frequencies of $\nu_{\rm beat} = 176$ and 214 cm$^{-1}$, respectively. The two peaks are found at an excitation energy of $E_{\rm exc} \approx 15650$ cm$^{-1}$. The detection energies of the two peaks are $E_{\rm det}$ = 15000 cm$^{-1}$ and 15580 cm$^{-1}$, which are the off-diagonal and diagonal peaks, respectively. The 2D maps obtained from the real part of the complex spectra are shown in Figs. 7(e) and (f). At a beat frequency of $\nu_{\rm beat} = 176$ cm$^{-1}$, the amplitudes of the peaks at $E_{\rm det}$ = 15000 cm$^{-1}$ and 15580 cm$^{-1}$ are negative and positive, respectively, which indicates that the phases of the quantum beats are opposite at these two detection energies. This feature can be assigned to the coherent oscillatory motions of a vibrational or vibronic wave packet on the exciton state during relaxation to the bottom of the potential energy surface.[20] The wave packet reflects from both sides of the potential energy surface, which are detected as 15000 $cm^{-1}$ and 15580 $cm^{-1}$. The amplitude of the quantum beat with a frequency of $\nu_{\text{beat}} = 214$ $cm^{-1}$ is opposite to that with $\nu_{\text{beat}} = 176$ $cm^{-1}$. The motion of the wave packet is illustrated schematically in Fig. 8(a) with a single vibrational coordinate. Although the one-dimensional potential energy curve is shown in Fig. 8, the two vibrational coordinates with these two frequencies are orthogonal to each other. The wave packet with a beat frequency of $\nu_{\text{beat}} = 176$ $cm^{-1}$ is first generated at the left wall of the potential. The wave packet then starts to move to the right and reflects from the right wall. Meanwhile, the wave packet with $\nu_{\text{beat}} = 214$ $cm^{-1}$ starts from the right wall and moves to the left.

Figures 7(c) and (d) show 2D coherent maps of CPC1 in intensity at beat frequencies of $\nu_{\text{beat}} = 202$ and 266 $cm^{-1}$, respectively. The three peaks are clearly seen in the map of CPC1 at an excitation energy of $E_{\text{exc}} \approx 16180$ $cm^{-1}$ which is close to the peak of the absorption spectra. The detection energies of the three peaks are $E_{\text{det}}$ = 15540 $cm^{-1}$, 15920 $cm^{-1}$, and 16320 $cm^{-1}$ with intervals of approximately 400 $cm^{-1}$. In contrast to APC, the amplitudes of all three peaks are negative in the maps obtained from the real part of the complex spectra, which suggests that the quantum beat oscillations of all three peaks are in phase. This cannot be explained by a simple picture of the wave packet motion as in the case of APC. The vibrational mode related to the vibrational wave packet of CPC1 may interact with the vibrations of other normal modes or the optically dark electronic states, which may be a reason why the coherence signature in CPC is weaker than that in APC.

As discussed in this section, the observed quantum beats originate from the vibrational coherence. It is possible that the low-frequency mode observed in the beat-frequency spectra is the mode promoting energy transfer. Low-frequency coherent vibronic motion with 200 $cm^{-1}$ has been observed not only in phycobiliproteins but also in other chromoproteins. However, this mode has not been considered to be an important vibrational mode for the relaxation process because this is in-plane isotropic motion that barely induces geometrical change through the vibration. However, the amplitude of this mode is strong, and the decay time of the quantum beat signal (~200 fs) is very close to the relaxation time of the vibrational energy on the lower exciton state.  It appears necessary to assign this coherent motion and to investigate whether this mode governs the relaxation process.

Furthermore, we have to assign whether the observed vibrational coherence is generated and detected on the electronically excited state *via* the stimulated emission (Fig. 8(b)) or on the electronic ground state *via* the ground state bleaching (Fig. 8(c)). In the previous studies of the Fenna-Matthews-Olson (FMO) complex, the observed vibrational coherence was assigned to that generated on the electronic ground state from the measurement of the mutant FMOs and independency with respect to the excitation energy.[52, 53] The coherence in the FMOs survived during long time after completion of the vibrational energy relaxation on the excited electronic state. On the other hand, the decay time of the vibrational coherence in the APC and CPC is similar to the time constant of the vibrational energy relaxation on the excited electronic state. From this similarity, we assign that the vibrational coherence on APC and CPC is generated on the electronically excited state.

## IV. SUMMARY AND PERSPECTIVES

Both chromoproteins, allophycocyanin (APC) and C-phycocyanin (CPC), have adjacent pairs of pigments α84 and β84, which are excited to the exciton states. However, we have revealed that their kinetics and coherence are significantly different. When APC is excited to the lower exciton state, the vibronic wave packet falls to the bottom of the potential energy surface with a time constant of 62 fs, while oscillating along the vibrational coordinate with a vibrational frequency of 200 $cm^{-1}$. The peak of the spectrum detected at the long delay time $T$ = 2 ps is shifted to the lower energy than the absorption spectrum, and narrower than that. These findings indicate that the population rapidly converges to the bottom of the potential energy surface. On the other hand, in CPC1, vibrational energy relaxation is not completed within 2 ps. The spectrum at $T$ = 2 ps is much broader than that of APC.

We found that proteins located on different positions in the phycobilisome possess different time constants and spectra, even for the same protein species. This should be one of the key mechanisms for the efficient and unidirectional energy transfer in the phycobilisome. To further investigate the intermolecular energy transfer between the protein molecules, the subsystems of the phycobilisome, for example, the complexes of CPC1–CPC2, CPC2–APC, and so on, will be synthesized artificially through genetic recombination, and the kinetics and coherence will be measured for each subsystem. When the new signatures that could not be observed in the constituent proteins are found in the subsystem, they can be identified as the interprotein energy transfer separated from the intraprotein energy relaxation.

## SUPPLEMENTARY MATERIAL

Refer to the supplementary material for detailed information on sample preparation, experimental apparatus for 2D-ES, and the additional dataset.

## ACKNOWLEDGMENTS

We thank Dr. Y. Yonetani and Dr. A. Tanaka at QST for valuable discussions of quantum coherence within biomolecules. We also thank Prof. A. Ishizaki at The University of Tokyo for valuable comments about 2D-ES. This research was supported by the MEXT Quantum Leap Flagship Program (JPMXS0120330644). We are grateful for the financial support of JSPS KAKENHI (JP21H01898) and the QST President's Strategic Grant (Exploratory Research). The mass spectrometry analysis shown in the Supplementary Materials was supported by J-PARC MLF deuteration laboratory. We thank Prof. K. Iwasaki at University of Tsukuba for valuable advices about the transmission electron microscopy analysis shown in the Supplementary Materials which was supported by Organization for Open Facility Initiatives in University of Tsukuba and Platform Project for Supporting Drug Discovery and Life Science Research (Basis for Supporting Innovative Drug Discovery and Life Science Research (BINDS)) from AMED under Grant Numbers JP24ama121001.

## AUTHOR DECLARATIONS

### Conflict of Interest

The authors have no conflicts to disclose.

### Author Contributions

**Masaaki Tsubouchi**: Conceptualization (equal); Data curation (equal); Formal analysis (lead); Funding acquisition (equal); Investigation (lead); Methodology (equal); Project administration (equal); Resources (equal); Software (lead); Supervision (equal); Visualization (lead); Writing – original draft (lead); Writing – review & editing (equal). **Nobuhisa Ishii**: Resources (equal); Writing – review & editing (equal). **Takatoshi Fujita**: Data curation (equal) ; Formal analysis (supporting); Methodology (equal); Writing – review & editing (equal). **Tomoyasu Adachi**: Conceptualization (equal); Funding acquisition (equal); Investigation (equal); Methodology (equal); Project administration (equal); Resources (equal); Supervision (equal); Writing – review & editing (equal). **Ryuji Itakura**: Project administration (equal); Supervision (equal); Writing – review & editing (equal).

## DATA AVAILABILITY

The data that support the findings of this study are available from the corresponding author upon reasonable request.

## REFERENCES

1. R. E. Blankenship, *Molecular mechanisms of photosynthesis*. (Wiley, 2021).

2. G. D. Scholes, G. R. Fleming, A. Olaya-Castro and R. van Grondelle, "Lessons from nature about solar light harvesting", Nat. Chem. **3**, 763-774 (2011).

3. G. S. Engel, T. R. Calhoun, E. L. Read, T.-K. Ahn, T. Mančal, Y.-C. Cheng, R. E. Blankenship and G. R. Fleming, "Evidence for wavelike energy transfer through quantum coherence in photosynthetic systems", Nature **446**, 782-786 (2007).

4. G. Panitchayangkoon, D. Hayes, K. A. Fransted, J. R. Caram, E. Harel, J. Wen, R. E. Blankenship and G. S. Engel, "Long-lived quantum coherence in photosynthetic complexes at physiological temperature", Proc. Natl. Acad. Sci. USA **107**, 12766-12770 (2010).

5. A. Ishizaki and G. R. Fleming, "Quantum coherence in photosynthetic light harvesting", Annu. Rev. Condens. Matter Phys. **3**, 333-361 (2012).

6. E. Collini, C. Y. Wong, K. E. Wilk, P. M. G. Curmi, P. Brumer and G. D. Scholes, "Coherently wired light-harvesting in photosynthetic marine algae at ambient temperature", Nature **463**, 644-647 (2010).

7. D. B. Turner, K. E. Wilk, P. M. G. Curmi and G. D. Scholes, "Comparison of electronic and vibrational coherence measured by two-dimensional electronic spectroscopy", J. Phys. Chem. Lett. **2**, 1904-1911 (2011).

8. D. B. Turner, R. Dinshaw, K.-K. Lee, M. S. Belsley, K. E. Wilk, P. M. G. Curmi and G. D. Scholes, "Quantitative investigations of quantum coherence for a light-harvesting protein at conditions simulating photosynthesis", Phys. Chem. Chem. Phys. **14**, 4857-4874 (2012).

9. G. H. Richards, K. E. Wilk, P. M. G. Curmi and J. A. Davis, "Disentangling electronic and vibrational coherence in the phycocyanin-645 light-harvesting complex", J. Phys. Chem. Lett. **5**, 43-49 (2014).

10. J. M. Womick and A. M. Moran, "Exciton coherence and energy transport in the light-harvesting dimers of allophycocyanin", J. Phys. Chem. B **113**, 15747-15759 (2009).

11. J. M. Womick, B. A. West, N. F. Scherer and A. M. Moran, "Vibronic effects in the spectroscopy and dynamics of *C*-phycocyanin", J. Phys. B: At. Mol. Opt. Phys. **45**, 154016 (2012).

12. H.-G. Duan, V. I. Prokhorenko, R. J. Cogdell, K. Ashraf, A. L. Stevens, M. Thorwart and R. J. D. Miller, "Nature does not rely on long-lived electronic quantum coherence for photosynthetic energy transfer", Proc. Natl. Acad. Sci. USA **114**, 8493-8498 (2017).

13. J. Cao, R. J. Cogdell, D. F. Coker, H.-G. Duan, J. Hauer, U. Kleinekathöfer, T. L. C.

Jansen, T. Mančal, R. J. D. Miller, J. P. Ogilvie, V. I. Prokhorenko, T. Renger, H.-S. Tan, R. Tempelaar, M. Thorwart, E. Thyrhaug, S. Westenhoff and D. Zigmantas, "Quantum biology revisited", Sci. Adv. **6**, eaaz4888 (2020).

14. T. Mirkovic, E. E. Ostroumov, J. M. Anna, R. van Grondelle, Govindjee and G. D. Scholes, "Light absorption and energy transfer in the antenna complexes of photosynthetic organisms", Chem. Rev. **117**, 249-293 (2017).

15. E. Gantt, "Phycobilisomes: light-harvesting pigment complexes", Bioscience **25**, 781-788 (1975).

16. L. Zheng, Z. Zheng, X. Li, G. Wang, K. Zhang, P. Wei, J. Zhao and N. Gao, "Structural insight into the mechanism of energy transfer in cyanobacterial phycobilisomes", Nat. Commun. **12**, 5497 (2021).

17. J. Ma, X. You, S. Sun, X. Wang, S. Qin and S. F. Sui, "Structural basis of energy transfer in porphyridium purpureum phycobilisome", Nature **579**, 146 (2020).

18. M. A. Domínguez-Martín, P. V. Sauer, H. Kirst, M. Sutter, D. Bína, B. J. Greber, E. Nogales, T. Polívka and C. A. Kerfeld, "Structures of a phycobilisome in light-harvesting and photoprotected states", Nature **609**, 835 (2022).

19. K. Kawakami, T. Hamaguchi, Y. Hirose, D. Kosumi, M. Miyata, N. Kamiya and K. Yonekura, "Core and rod structures of a thermophilic cyanobacterial light-harvesting phycobilisome", Nat. Commun. **13**, 3389 (2022).

20. S. Sil, R. W. Tilluck, T. M. N. Mohan, C. H. Leslie, J. B. Rose, M. A. Domínguez-Martín, W. Lou, C. A. Kerfeld and W. F. Beck, "Excitation energy transfer and vibronic coherence in intact phycobilisomes", Nat. Chem. **14**, 1286 (2022).

21. S. Sohoni, L. T. Lloyd, A. Hitchcock, C. MacGregor-Chatwin, A. Iwanicki, I. Ghosh, Q. Shen, C. N. Hunter and G. S. Engel, "Phycobilisome's exciton transfer efficiency relies on an energetic funnel driven by chromophore–linker protein interactions", J. Am. Chem. Soc. **145**, 11659-11668 (2023).

22. P. Navotnaya, S. Sohoni, L. T. Lloyd, S. M. Abdulhadi, P. C. Ting, J. S. Higgins and G. S. Engel, "Annihilation of excess excitations along phycocyanin rods precedes downhill flow to allophycocyanin cores in the phycobilisome of synechococcus elongatus PCC 7942", J. Phys. Chem. B **126**, 23 (2022).

23. Y. Hirota, H. Serikawa, K. Kawakami, M. Ueno, N. Kamiya and D. Kosumi, "Ultrafast energy transfer dynamics of phycobilisome from thermosynechococcus vulcanus, as revealed by ps fluorescence and fs pump-probe spectroscopies", Photosynth. Res. **148**, 181 (2021).

24. K. Brejc, R. Ficner, R. Huber and S. Steinbacher, "Isolation, crystallization, crystal structure analysis and refinement of allophycocyanin from the cyanobacterium Spirulina platensis at 2.3 Å resolution", J. Mol. Biol. **249**, 424-440 (1995).

25. R. Zhu, W. Li, Z. Zhen, J. Zou, G. Liao, J. Wang, Z. Wang, H. Chen, S. Qin and Y. Weng, "Quantum phase synchronization via exciton-vibrational energy dissipation sustains long-lived coherence in photosynthetic antennas", Nat. Commun. **15**, 3171 (2024).

26. R. MacColl, "Allophycocyanin and energy transfer", Biochim. Biophys. Acta **1657**, 73-81 (2004).

27. M. D. Edington, R. E. Riter and W. F. Beck, "Evidence for coherent energy transfer in allophycocyanin trimers", J. Phys. Chem. **99**, 15699-15704 (1995).

28. M. D. Edington, R. E. Riter and W. F. Beck, "Interexciton-state relaxation and exciton localization in allophycocyanin trimers", J. Phys. Chem. B **100**, 14206-14217 (1996).

29. M. D. Edington, R. E. Riter and W. F. Beck, "Femtosecond transient hole-burning detection of interexciton-state radiationless decay in allophycocyanin rrimers", J. Phys. Chem. B **101**, 4473-4477 (1997).

30. B. J. Homoelle, M. D. Edington, W. M. Diffey and W. F. Beck, "Stimulated photon-echo and transient-grating studies of protein-matrix solvation dynamics and interexciton-state radiationless decay in α phycocyanin and allophycocyanin", J. Phys. Chem. B **102**, 3044-3052 (1998).

31. J. M. Zhang, Y. J. Shiu, M. Hayashi, K. K. Liang, C. H. Chang, V. Gulbinas, C. M. Yang, T. S. Yang, H. Z. Wang, Y. T. Chen and S. H. Lin, "Investigations of ultrafast exciton dynamics in allophycocyanin trimer", J. Phys. Chem. A **105**, 8878-8891 (2001).

32. J. M. Womick and A. M. Moran, "Nature of excited states and relaxation mechanisms in C-phycocyanin", J. Phys. Chem. B **113**, 15771 (2009).

33. J. M. Womick, S. A. Miller and A. M. Moran, "Toward the origin of exciton electronic structure in phycobiliproteins", J. Chem. Phys. **133**, 024507 (2010).

34. J. M. Womick and A. M. Moran, "Vibronic enhancement of exciton sizes and energy transport in photosynthetic complexes", J. Phys. Chem. B **115**, 1347-1356 (2011).

35. A. Fălămaş, S. A. Porav and V. Tosa, "Investigations of the energy transfer in the phycobilisome antenna of arthrospira platensis using femtosecond spectroscopy", Appl. Sci. **10**, 4045 (2020).

36. S. Mukamel, "Multidimensional femtosecond correlation spectroscopies of electronic and vibrational excitations", Annu. Rev. Phys. Chem. **51**, 691-729 (2000).

37. J. D. Hybl, A. A. Ferro and D. M. Jonas, "Two-dimensional Fourier transform electronic spectroscopy", J. Chem. Phys. **115**, 6606-6622 (2001).

38. D. M. Jonas, "Two-dimensional femtosecond spectroscopy", Annu. Rev. Phys. Chem. **54**, 425-463 (2003).

39. T. Mančal, A. V. Pisliakov and G. R. Fleming, "Two-dimensional optical three-pulse photon echo spectroscopy. I. Nonperturbative approach to the calculation of spectra", J. Chem. Phys. **124**, 234504 (2006).

40. A. V. Pisliakov, T. Mančal and G. R. Fleming, "Two-dimensional optical three-pulse photon echo spectroscopy. II. Signatures of coherent electronic motion and exciton population transfer in dimer two-dimensional spectra", J. Chem. Phys. **124**, 234505 (2006).

41. T. A. A. Oliver, "Recent advances in multidimensional ultrafast spectroscopy", R. Soc. Open Sci. **5**, 171425 (2018).

42. A. Gelzinis, R. Augulis, V. Butkus, B. Robert and L. Valkunas, "Two-dimensional spectroscopy for non-specialists", Biochim. Biophys. Acta **1860**, 271-285 (2019).

43. E. Romero, V. I. Novoderezhkin and R. van Grondelle, "Quantum design of photosynthesis for bio-inspired solar-energy conversion", Nature **543**, 355-365 (2017).

44. G. D. Scholes, G. R. Fleming, L. X. Chen, A. Aspuru-Guzik, A. Buchleitner, D. F. Coker, G. S. Engel, R. van Grondelle, A. Ishizaki, D. M. Jonas, J. S. Lundeen, J. K. McCusker, S. Mukamel, J. P. Ogilvie, A. Olaya-Castro, M. A. Ratner, F. C. Spano, K. B. Whaley and X. Zhu, "Using coherence to enhance function in chemical and biophysical systems", Nature **543**, 647-656 (2017).

45. M. Tsubouchi, N. Ishii, Y. Kagotani, R. Shimizu, T. Fujita, M. Adachi and R. Itakura, "Beat-frequency-resolved two-dimensional electronic spectroscopy: disentangling vibrational coherences in artificial fluorescent proteins with sub-10-fs visible laser pulses", Opt. Express **31**, 6890-6906 (2023).

46. S.-H. Shim and M. T. Zanni, "How to turn your pump–probe instrument into a multidimensional spectrometer: 2D IR and Vis spectroscopiesvia pulse shaping", Phys. Chem. Chem. Phys. **11**, 748-761 (2009).

47. D. Brida, C. Manzoni and G. Cerullo, "Phase-locked pulses for two-dimensional spectroscopy by a birefringent delay line", Opt. Lett. **37**, 3027-3029 (2012).

48. J. Réhault, M. Maiuri, A. Oriana and G. Cerullo, "Two-dimensional electronic spectroscopy with birefringent wedges", Rev. Sci. Instrum. **85**, 123107 (2014).

49. V. R. Policht, A. Niedringhaus, R. Willow, P. D. Laible, D. F. Bocian, C. Kirmaier, D.

Holten, T. Mančal and J. P. Ogilvie, "Hidden vibronic and excitonic structure and vibronic coherence transfer in the bacterial reaction center", Sci. Adv. **8**, eabk0953 (2022).

50. F. D. Fuller, J. Pan, A. Gelzinis, V. Butkus, S. S. Senlik, D. E. Wilcox, C. F. Yocum, L. Valkunas, D. Abramavicius and J. P. Ogilvie, "Vibronic coherence in oxygenic photosynthesis", Nat. Chem. **6**, 706-711 (2014).

51. E. Romero, R. Augulis, V. I. Novoderezhkin, M. Ferretti, J. Thieme, D. Zigmantas and R. van Grondelle, "Quantum coherence in photosynthesis for efficient solar-energy conversion", Nat. Phys. **10**, 676-682 (2014).

52. M. Maiuri, E. E. Ostroumov, R. G. Saer, R. E. Blankenship and G. D. Scholes, "Coherent wavepackets in the Fenna–Matthews–Olson complex are robust to excitonic-structure perturbations caused by mutagenesis", Nature Chem. **10**, 177-183 (2018).

53. E. Thyrhaug, R. Tempelaar, M. J. P. Alcocer, K. Žídek, D. Bína, J. Knoester, T. L. C. Jansen and D. Zigmantas, "Identification and characterization of diverse coherences in the Fenna–Matthews–Olson complex", Nature Chem. **10**, 780-786 (2018).

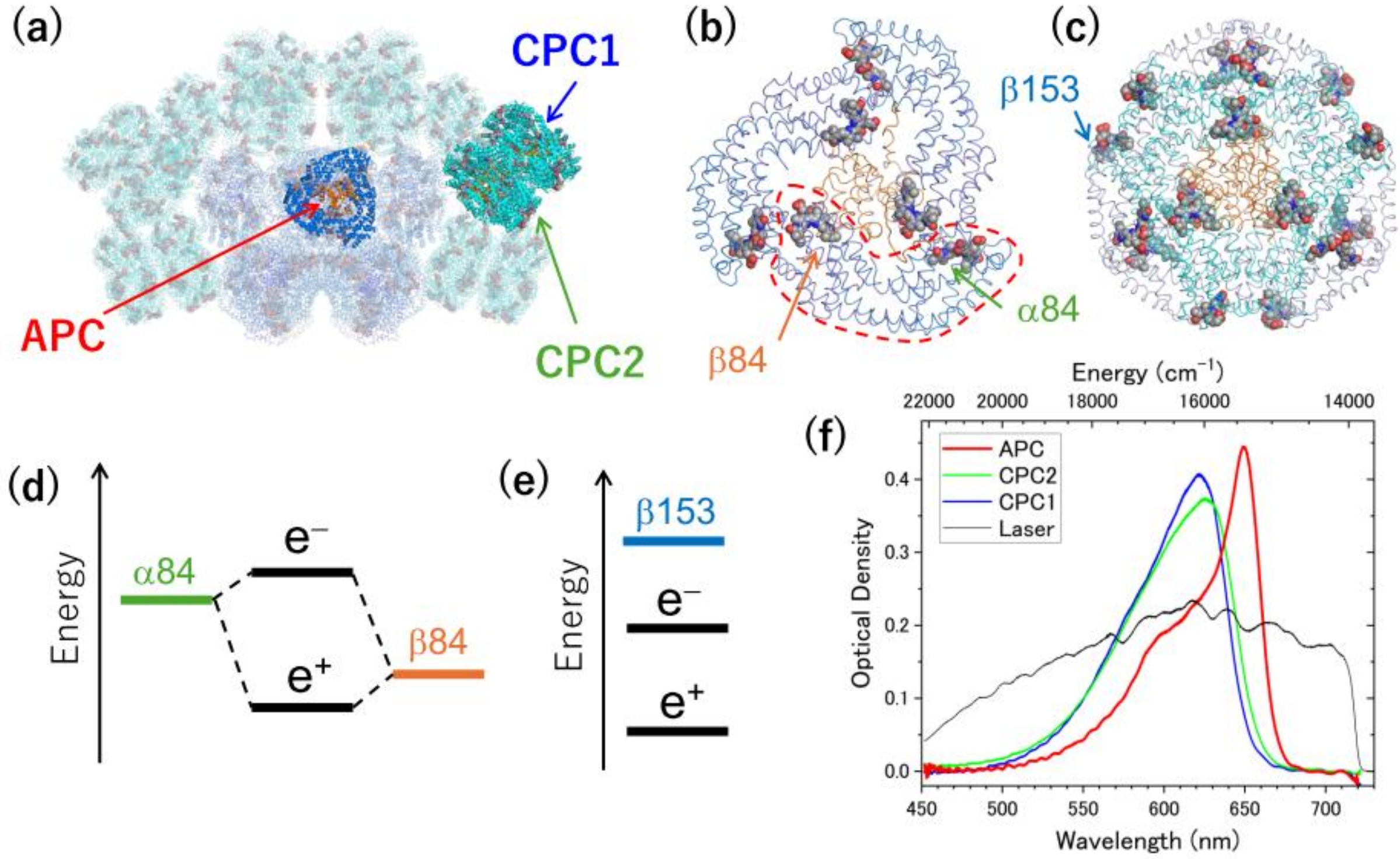

**FIG. 1.** (a) Cryo-EM structure of *Anabaena* sp. PCC 7120 phycobilisome referred to in this study. The APC trimer and two kinds of CPC hexamers are highlighted.[19] (b) Structure of the APC trimer in which the α84 and β84 pigments are shown with a van der Waals representation. The APC monomer unit is indicated by the red dashed line. (c) Structure of the CPC hexamer. In addition to the α84 and β84 pigments, which are covalently bonded to the polypeptide in similar positions to the APC, the β153 pigments are bounded at the outermost position. (d) Energy level diagram of APC based on the purely electronic exciton model. (e) Energy level diagram of CPC. (f) Absorption spectra of APC (red) and the two kinds of CPCs (green and blue) from *Thermosynechococcus elongatus* BP-1 phycobilisome. The broadband spectrum of the excitation and detection laser pulses is shown as the thin black line.

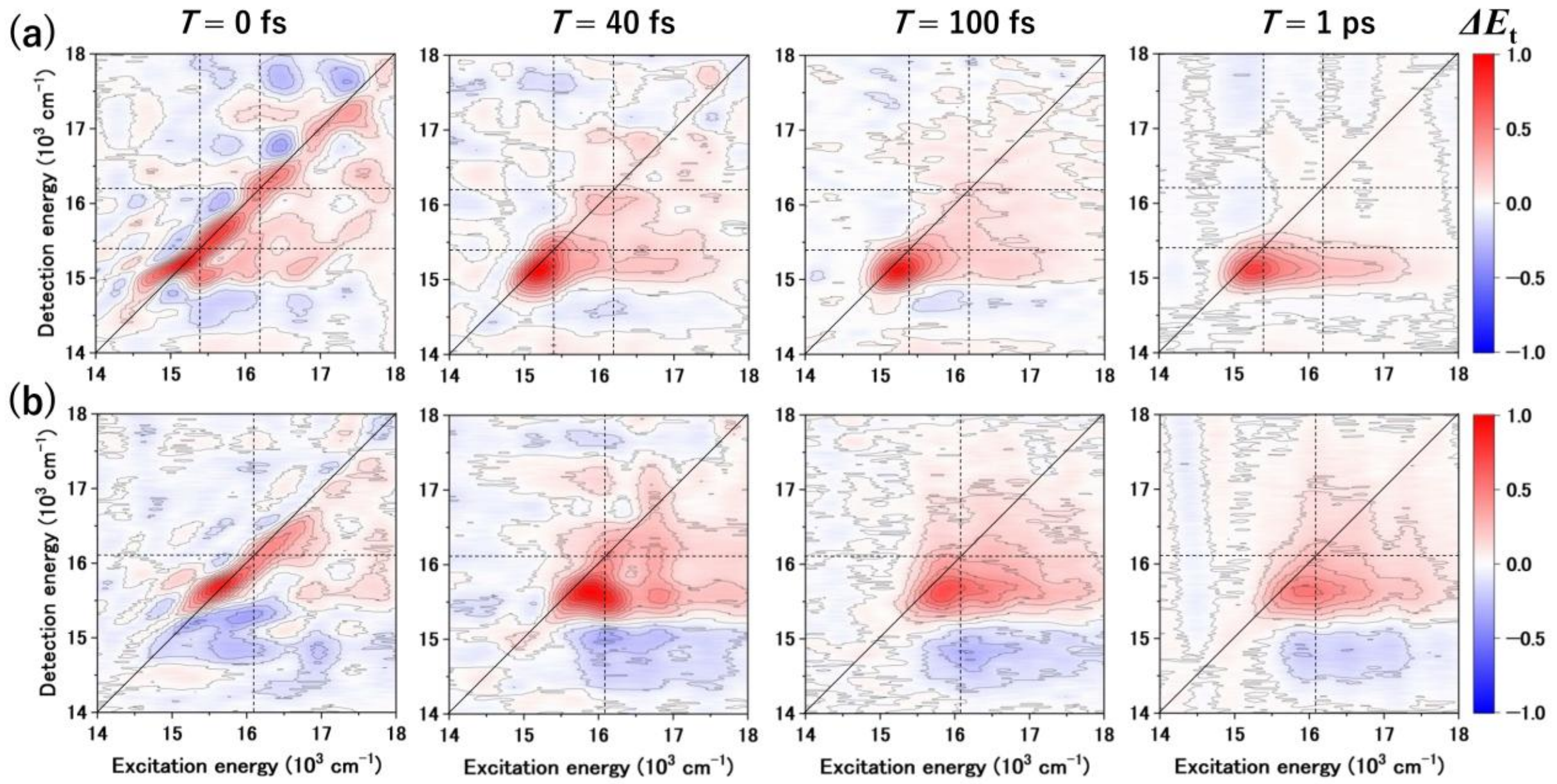

**FIG. 2.** (a) A series of absorptive real-valued 2D spectra of APC measured at pump–probe delay times of $T$ = 0 fs, 40 fs, 100 fs, and 1 ps. The positive amplitude represents increasing transmission, which indicates GSB or SE, while the negative amplitude indicates ESA. The dashed lines show the energy levels of the higher (16200 cm$^{-1}$) and lower (15400 cm$^{-1}$) exciton states shown in Fig. 1(f). (b) Absorptive real-valued 2D spectra of CPC1. The dashed lines show the peak energy (16100 cm$^{-1}$) of the absorption spectrum shown in Fig. 1(f).

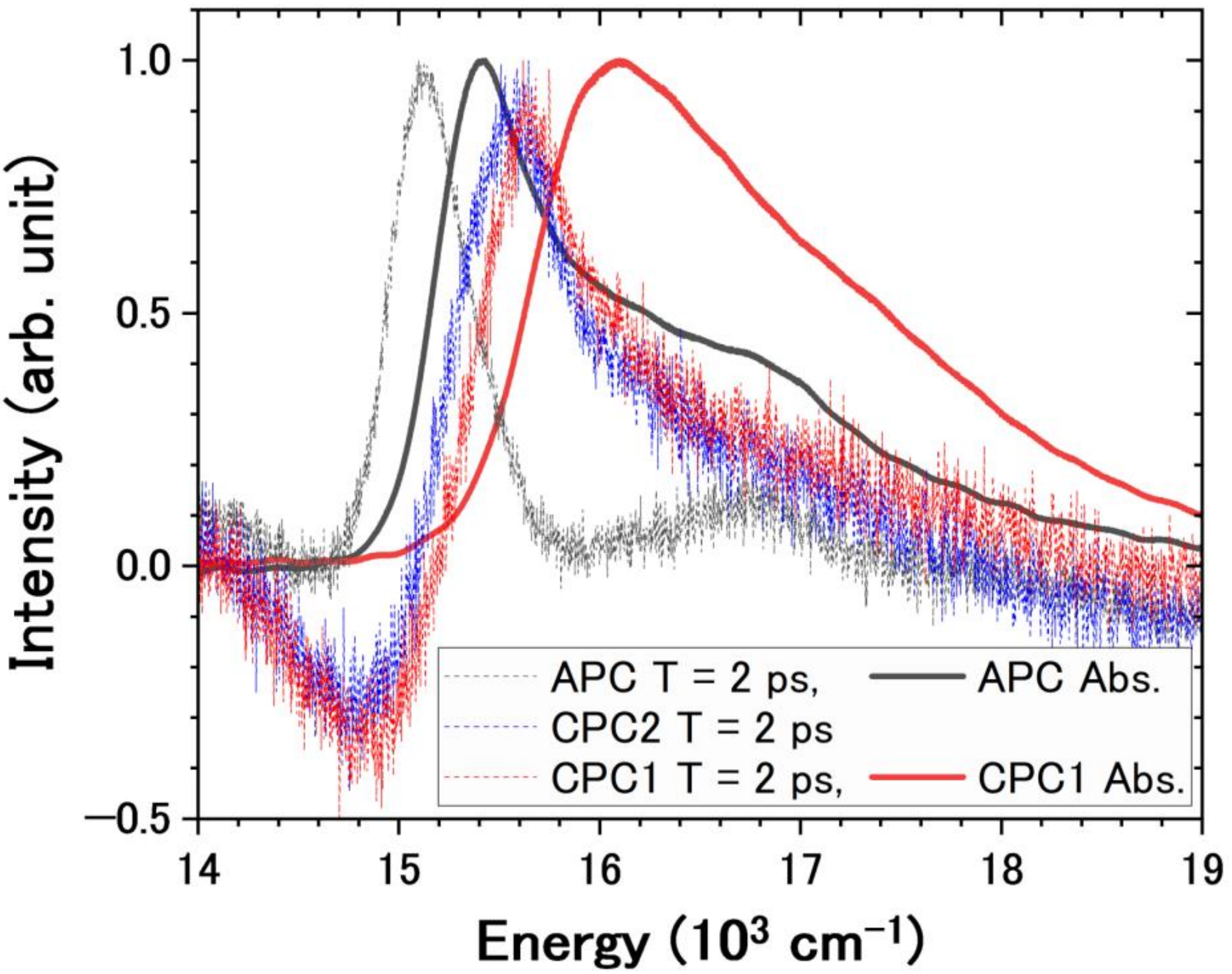

**FIG. 3.** Comparison between the transient spectra at the long delay time ($T$ = 2 ps) and the absorption spectra. The thin dashed lines are the transient spectra obtained by projection of the 2D electronic spectra onto the detection energy axis. The thick solid lines are the absorption spectra. The black, blue, and red lines represent the spectra of APC, CPC2, and CPC1, respectively. The intensities are adjusted such that the maximum value is 1.

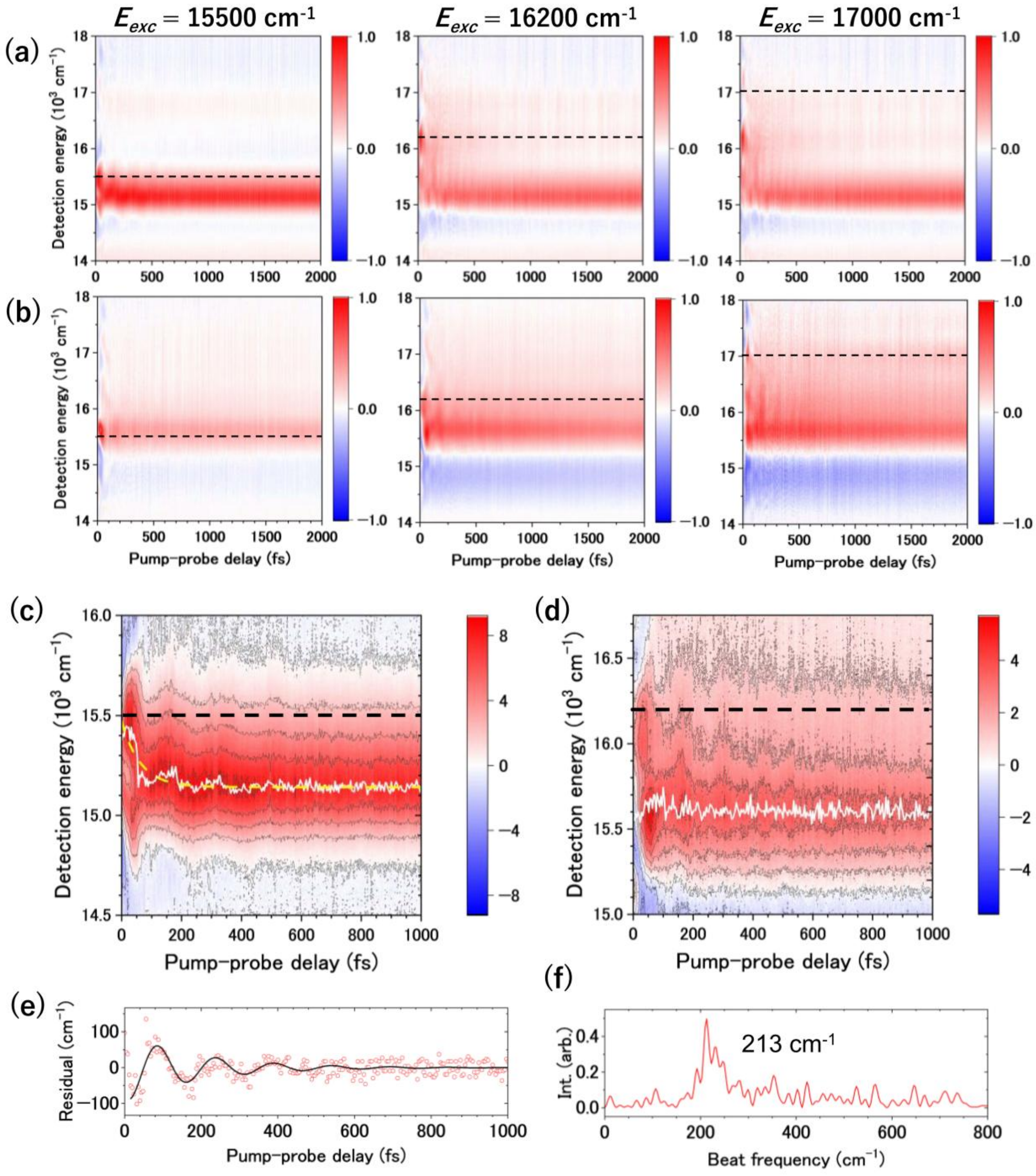

**FIG. 4.** Time-resolved detection signal spectra $S(E_{\mathrm{det}}, T\colon E_{\mathrm{exc}})$ of (a) APC and (b) CPC1 at excitation energies of $E_{\mathrm{exc}}$ = 15500, 16200, and 17000 cm$^{-1}$. The dashed lines indicate the excitation energies. Enlarged views of $S(E_{\mathrm{det}}, T\colon E_{\mathrm{exc}})$ are shown in (c) and (d) for APC at $E_{\mathrm{exc}}$ = 15500 cm$^{-1}$ and CPC1 at $E_{\mathrm{exc}}$ = 16200 cm$^{-1}$, respectively. The dashed horizontal lines mark the excitation energies. The white solid lines show the time profile of the detection energy with the maximum amplitude. The yellow dashed curve shows the fitting result to the observed time profile with a single exponential decay function. (e) Oscillatory residual curve of APC at $E_{\mathrm{exc}}$ = 15500 cm$^{-1}$ obtained from the exponential fitting in (c). Red open dots are the residues. The black solid line is the fitting result with the damped oscillation $\exp(-T/\tau_c)\sin\{2\pi\nu(T-T_D)\}$, where $\tau_c$ is the time constant of decoherence, $\nu$ is the frequency of the beat, and $T_D$ is the phase delay of the beat. (f) Fourier transform spectrum of the residual time profile.

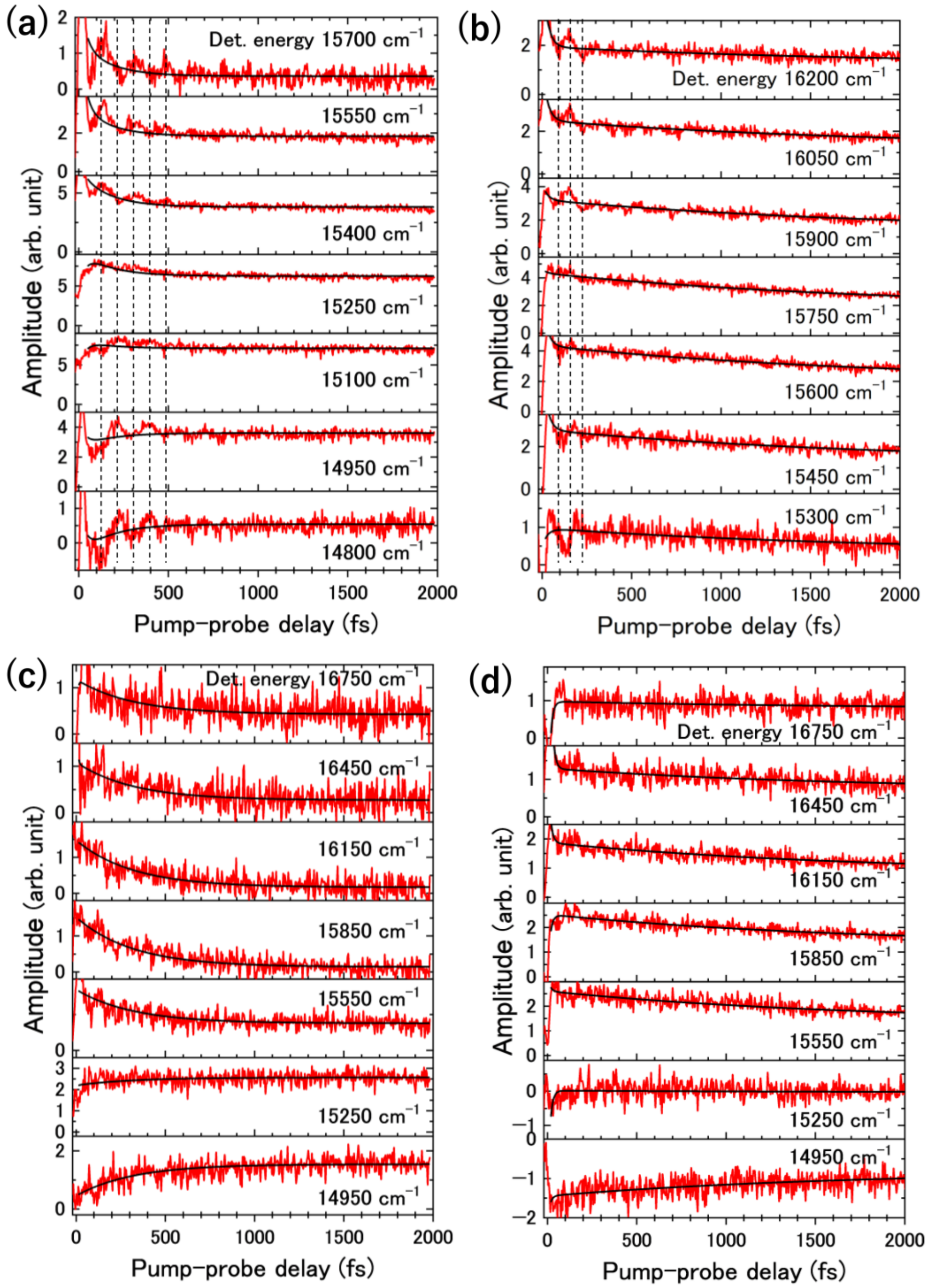

**FIG. 5.** (a) Time profiles of the signal amplitude after excitation measured for APC. The excitation energy is the peak of the absorption spectra, 15500 $cm^{-1}$. The red lines show the experimental data. The black lines show the result fitted by the model function Eq. (1) using time constants, $\tau_f$ and $\tau_s$, obtained by global analysis. The detection energies are indicated close to the profiles. The vertical dashed lines indicate the peaks and valleys. (b) Time profiles of CPC1 measured at the excitation energy of 16150 $cm^{-1}$. (c) Time profiles of APC measured at the excitation energy of 16650 $cm^{-1}$. (d) Time profiles of CPC1 measured at the excitation energy of 16650 $cm^{-1}$.

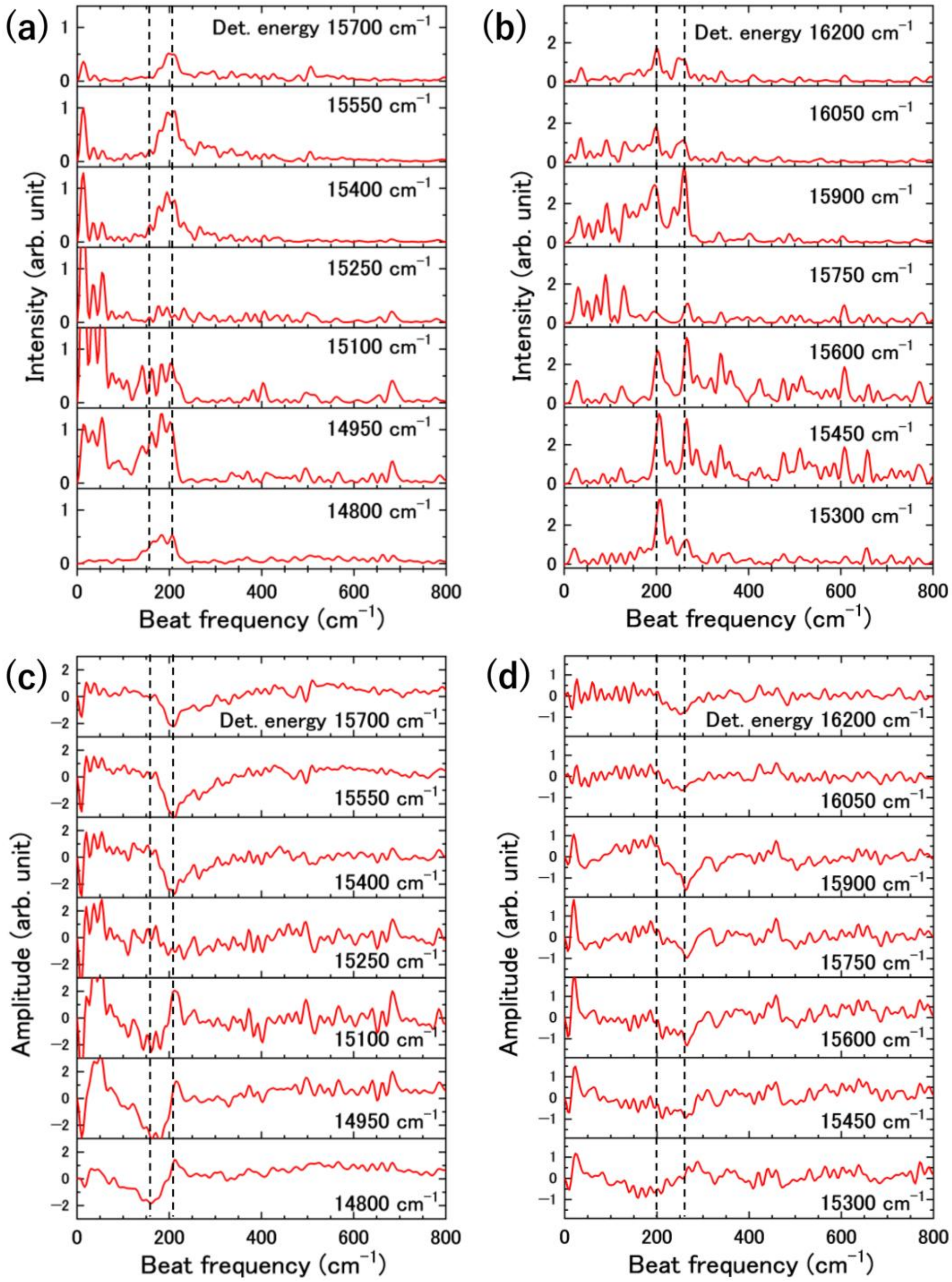

**FIG. 6.** Beat-frequency spectra calculated from the Fourier transform of the residual time profiles. (a) and (b) are the intensity spectra of APC at an excitation energy of $E_{\mathrm{exc}} = 15500$ cm$^{-1}$ and of CPC1 at $E_{\mathrm{exc}} = 16150$ cm$^{-1}$, respectively. (c) and (d) are the amplitude spectra (the real part of the complex Fourier transform spectra) corresponding to the intensity spectra of (a) and (b), respectively. The detection energies are indicated close to the profiles.

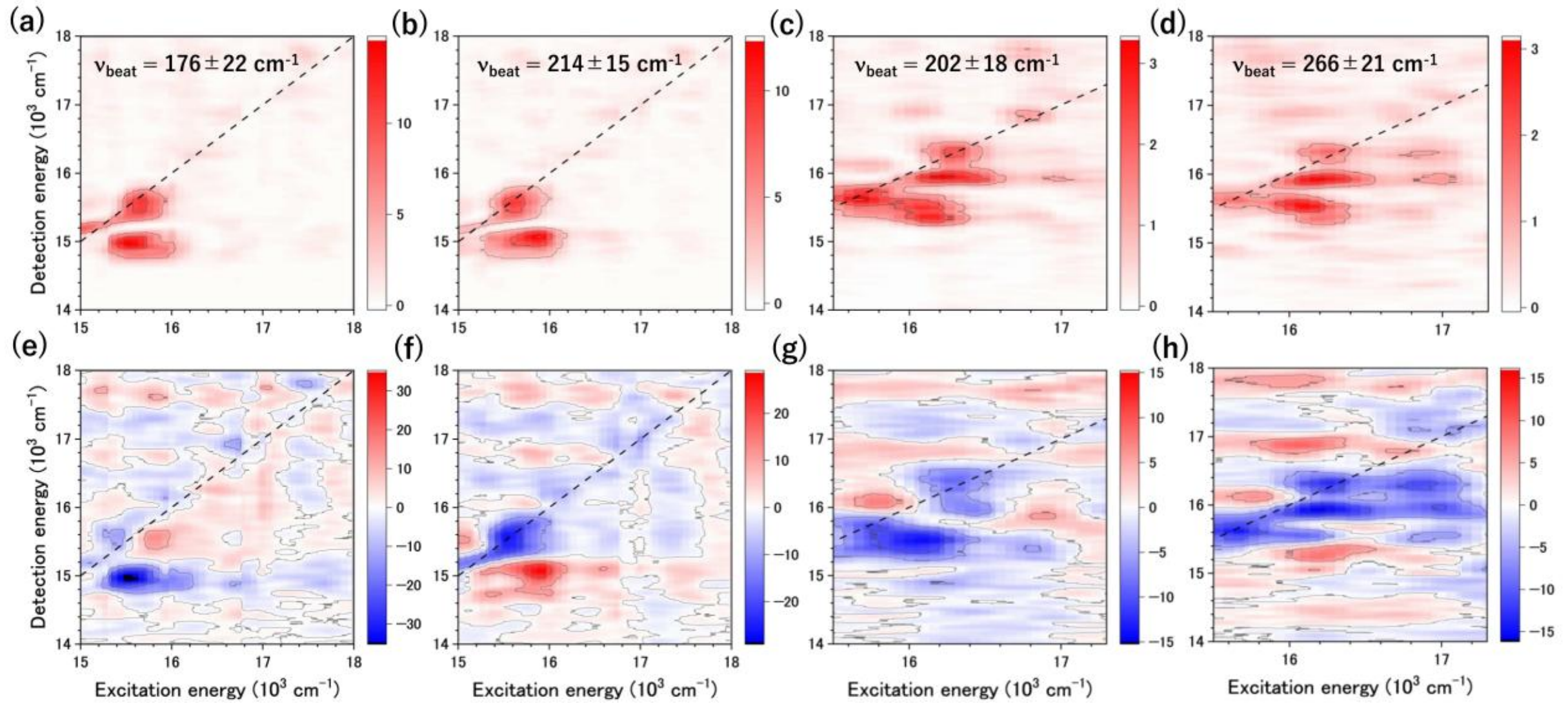

**FIG. 7.** Beat-frequency-resolved 2D electronic spectra. (a) 2D spectrum of APC calculated at the beat frequency $\nu_{\text{beat}}$ = 176 ± 22 cm$^{-1}$. The dashed line indicates the diagonal. (b) 2D spectrum of APC at $\nu_{\text{beat}}$ = 214 ± 15 cm$^{-1}$. (c) 2D spectrum of CPC1 at $\nu_{\text{beat}}$ = 202 ± 18 cm$^{-1}$. (d) 2D spectrum of CPC1 at $\nu_{\text{beat}}$ = 266 ± 21 cm$^{-1}$. (a), (b), (c), and (d) show the 2D spectra in intensity. (e), (f), (g), and (h) are the amplitude 2D spectra (the real part of the complex Fourier transform spectra) corresponding to the intensity spectra of (a), (b), (c), and (d), respectively.

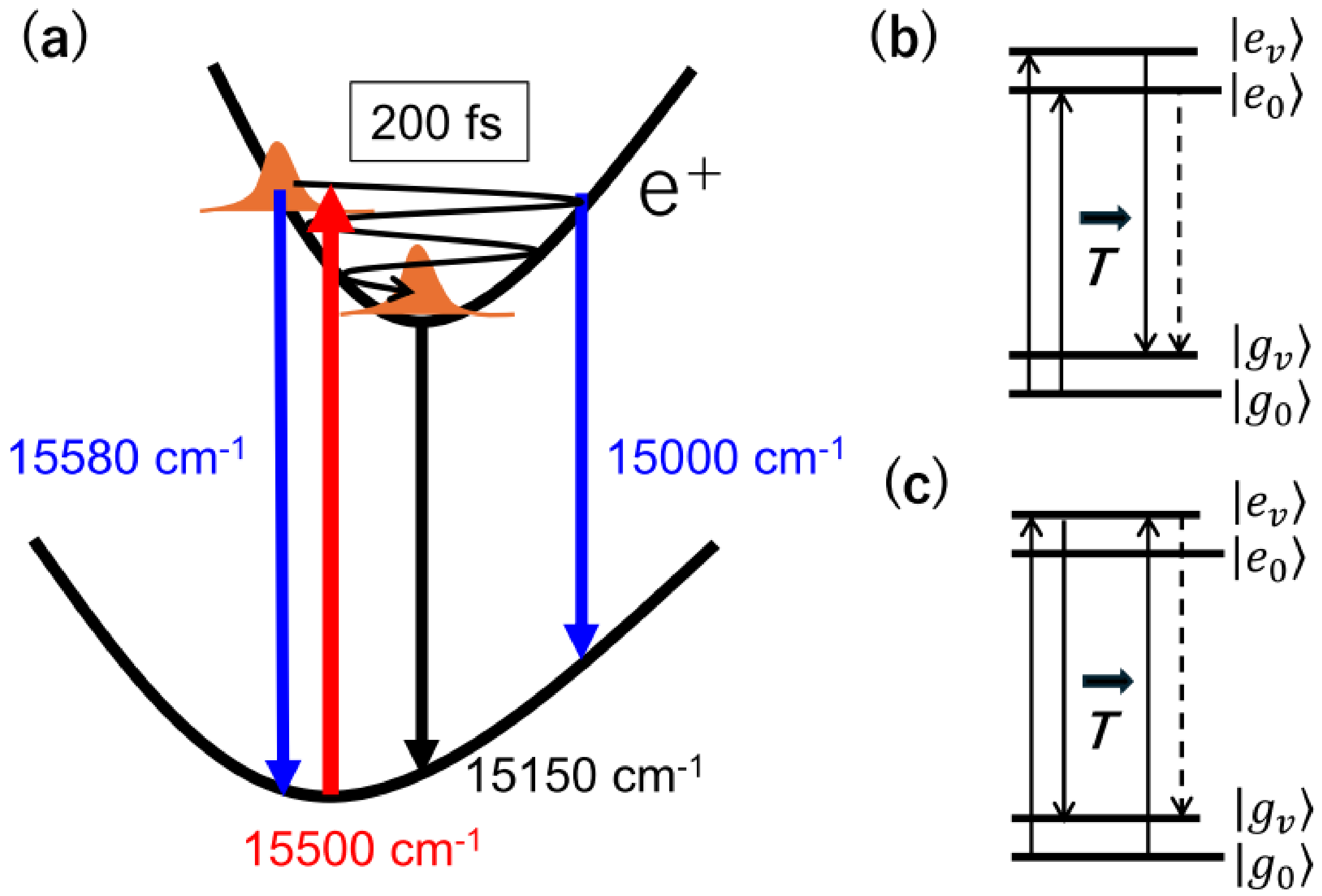

**FIG. 8.** (a) Vibrational wave packet generated on the lower exciton state. The numbers indicate the vertical energies in the excitation (red) and detection (blue) processes. (b) and (c) shows one of the pathways of the four-wave mixing in SE and GSB processes, respectively. The solid arrows show the pump and probe laser light, and the dashed arrow shows the detected signal light. $|g_0\rangle$ and $|g_v\rangle$ represent a ground and an excited vibrational states on the electronic ground state, respectively. $|e_0\rangle$ and $|e_v\rangle$ represent a ground and an excited vibronic states on the excitonic state, respectively.

Supplemental Issues

# Dependence of energy relaxation and vibrational coherence on the location of light-harvesting chromoproteins in photosynthetic antenna protein complexes

**Masaaki Tsubouchi,[1,2,a)] Nobuhisa Ishii,[1] Takatoshi Fujita,[2] Motoyasu Adachi,[2] and Ryuji Itakura[1]**

**AFFILIATIONS**

[1]Kansai Institute for Photon Science (KPSI), National Institutes for Quantum Science and Technology (QST), 8-1-7 Umemidai, Kizugawa, Kyoto 619-0215, Japan

[2]Institute for Quantum Life Science, National Institutes for Quantum Science and Technology (QST), 4-9-1 Anagawa, Inage, Chiba 263-8555, Japan

[a)]Author to whom correspondence should be addressed: tsubouchi.masaaki@qst.go.jp

**Contents**

## 1. Sample preparation

The three kinds of recombinant proteins (APC, CPC1, and CPC2 from *Thermosynechococcus elongatus* BP-1 shown in Fig. 1(a) in the main text) were produced by an *Escherichia coli* (*E. coli*) expression system. Supplementary Table S1 and Fig. S1 list the expression plasmids used in this study. DNA and encoded amino acid sequences of proteins were deposited into the DNA Data Bank of Japan. The sequences were designed with a three-dimensional structure.[1-6] The molecules of C-phycocyanin (CPC) and allophycocyanin (APC) generally consist of alpha and beta chains and form a ring structure as a trimer or hexamer of the heterodimer (alpha and beta chains) in a large complex of phycobilisome. The alpha chain of CPC and alpha and beta chains of APC link one phycocyanobilin via a covalent bond as a chromophore in native phycobilisome, whereas the beta chains of CPC link two phycocyanobilins. ApcC and partial ApcE were co-expressed to prepare APC, while CpcC, partial CpcD, and partial CpcG2 were co-expressed to prepare CPC1 or CPC2 in order to reproduce the native structure. It is known that those five types of accessory molecules do not include any chromophores. All DNAs were chemically synthesized and purchased from Genewiz (Azenta Life Sciences, Japan), and were optimized for *E. coli*. The expression plasmids were transfected into *E. coli* JM109(DE3) (Promega). The pACYC_HO1PcyA and pCDF_MTSEF were used for posttranslational modifications of APC, CPC1, and CPC2.

The bacterial cells were grown at 310 K in Terrific Broth medium containing 15 mg/L kanamycin, 7 mg/L chloramphenicol, and 10 mg/L streptomycin to an OD600 of around 0.7. Then, protein expression was induced at 310 K for 48 hours by adding isopropyl-β-D-thiogalactopyranoside to a final concentration of 0.1 mM, as well as 5-aminolevulinic acid (final 80 mg/mL), biotin (final 0.25 mg/L), vitamin B12 (final 0.135 mg/L), and thiamine (final 0.335 mg/L). The cells were harvested by centrifugation at 5000 g for 20 min. The harvested cells from a 0.6 L of culture were suspended into 30 mL of Buffer T (20 mM Tris-HCl, pH 8.0) and disrupted by sonication. The crude extract was centrifuged at 10000 g for 30 min to collect the supernatant. The supernatant was dialyzed in 1 L of Buffer T overnight and then centrifuged at 10000 g for 30 min to remove the insoluble fraction. The obtained supernatant was applied onto a column filled with Ni Sepharose 6 Fast Flow (Cytiva) at 20 mL volume and eluted with 0.5 M imidazole in Buffer T. The eluted solution was applied onto a StrepTrap XT column with 5 mL volume (Cytiva) and eluted by biotin in Buffer T. Subsequently, a solution purified by StrepTrap was applied onto a HiTrap Q XL column (Cytiva). The protein was eluted with a linear gradient of 0–0.5 M NaCl in Buffer T. The pooled fraction containing protein was desalted with Buffer T by dialysis. The protein concentrations of APC, CPC1, and CPC2 were estimated using UV absorption at 280 nm with molecular extinction coefficients of $25.4 \times 10^7$, $26.3 \times 10^7$, and $26.1 \times 10^7$ $cm^{-1}$ $M^{-1}$, respectively, as calculated from the amino acid content.[7] For the measurement of two-dimensional (2D) electronic spectra, the protein solution included 10% w/v glycerol and 0.5 mM tris(2-carboxyethyl)phosphine hydrochloride in Buffer T.

**Supplementary Table S1** | List of expression plasmids used in this study.

| Name of plasmid | Original plasmid | Coded proteins | Organisms* | Accession number (reference sequence) | Accession number deposited |
|---|---|---|---|---|---|
| pET_TeAPC (C'E') [APC] | pET24a (Novagen) | ApcA | BP-1 | NCBI: WP_011056801 | LC853280 |
| | | ApcB | BP-1 | NCBI: WP_011056800 | |
| | | ApcC' | BP-1 | NCBI: WP_011056799 | |
| | | ApcE' | BP-1 | NCBI: WP_011058198 | |
| pET_TeCPC (D'C') [CPC1] | pET24a (Novagen) | CpcA | BP-1 | NCBI: WP_011057793 | LC853281 |
| | | CpcB | BP-1 | NCBI: WP_011057792 | |
| | | CpcD' | BP-1 | NCBI: WP_011057795 | |
| | | CpcC' | BP-1 | NCBI: WP_011058198 | |
| pET_TeCPC (G2'C') [CPC2] | pET24a (Novagen) | CpcA | BP-1 | NCBI: WP_011057793 | LC853282 |
| | | CpcB | BP-1 | NCBI: WP_011057792 | |
| | | CpcG2' | BP-1 | NCBI: WP_011057799 | |
| | | CpcC'[2] | BP-1 | NCBI: WP_011057794 | |
| pACYC_ HO1PcyA | pACYCDuet-1 (Novagen) | HO1 | 6803 | NCBI: WP_010871494 | LC853284 |
| | | PcyA | 7120 | NCBI: WP_010997850 | |
| pCDF_ MTSEF | pCDFDuet-1 (Novagen) | CpcM | BP-1 | NCBI: WP_011057783 | LC853283 |
| | | CpcT | 7120 | NCBI: WP_010999463 | |
| | | CpcS | 7120 | NCBI: WP_010994793 | |
| | | CpcE | 7120 | NCBI: WP_010994708 | |
| | | CpcF | 7120 | NCBI: WP_010994709 | |

*BP-1: *Thermosynechococcus elongatus* BP-1 (*Thermosynechococcus vestitus* BP-1)

6803: *Synechocystis* sp. PCC 6803

7120: *Nostoc* sp. 7120 (*Anabaena* sp.)

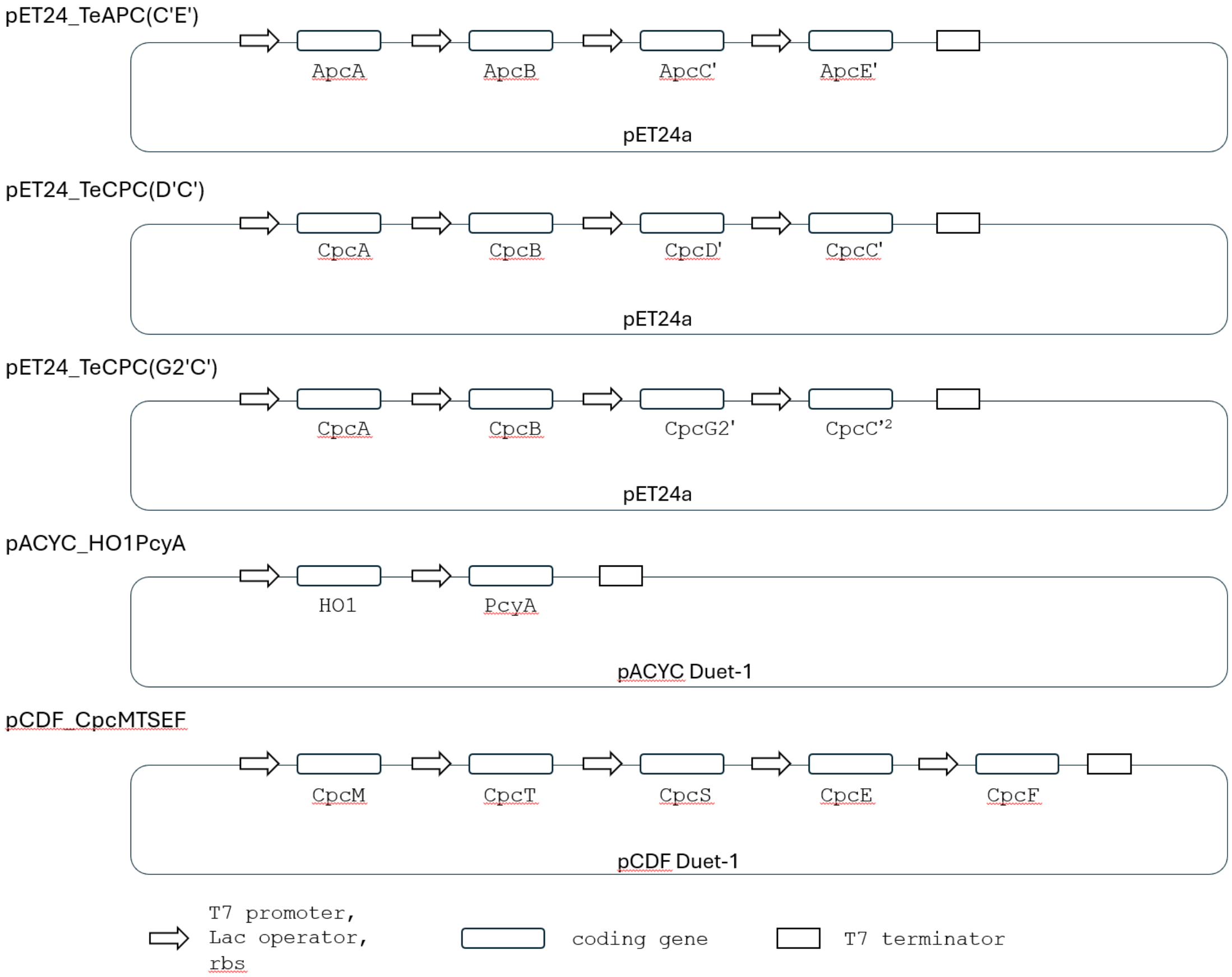

**Supplementary Fig. S1 | Constructed expression plasmids.** Constructions of the expression plasmids used in this study. pACYC_HO1PcyA was used for co-expression to synthesize pigments of phycocyanobilin in *E coli*. pCDF_MTSEF was used for co-expression to modify APC and CPC by encoded enzymes. In pET24a, synthesized DNA was inserted between NdeI and HindIII restriction enzyme sites. HO1, PcyA and CpcM were inserted into the MCS1 region, and the other synthesized genes were inserted into the MCS2 region.

## 2. Sample characterization

To confirm the polypeptides included in the purified protein, APC, CPC1, and CPC2 were analyzed by sodium dodecyl sulfate–polyacrylamide gel electrophoresis (SDS-PAGE). Supplementary Fig. S2 shows a photo of the stained gel. It appears that two polypeptide bands of ApcA and ApcB are combined because of their similar size. Stained bands of CpcC' and CpcG2' are not clearly observable, whereas the molecular ratio appears to be one-sixth that of CpcA and CpcB, which corresponds to the molecular ratio of ApcE' to ApcA and ApcB. Any bands of smaller molecular weight for ApcC', CpcD', and CpcC'[2] were not observed.

To further confirm the existence of polypeptides and posttranslational modification, proteins were analyzed by mass spectrometry. The purified protein samples were injected into an ultra-high-performance liquid chromatography system in tandem with a quadruple time-of-flight mass spectrometer (H-Class and Xevo G2-XS; Milford, MA, USA, Waters). The intact proteins were desalted by chromatographic separation on a MassPREP Micro Desalting Column (Waters) with a gradient from water to acetonitrile containing 0.1% formic acid. Furthermore, the proteins were fragmented by protease LysC and separated on an Acquity UPLC BEH C18 column (2.1 × 50 mm, Waters) for peptide mapping analysis. An electrospray ionization technique was employed in positive ion modes. Leucine enkephalin and NaI were used for the following quantifier ion and calibration, respectively. Raw data outputted by MassLynx were analyzed by UNIFI version 1.9. Mass analysis shows that phycocyanobilin is attached to the alpha and beta chains, as indicated by the natural proteins purified from cyanobacteria. Methylation at the beta chains suggests that CpcM has modified asparagine residues. Although ApcC', CpcD', and CpcC'2 were not observed in SDS-PAGE analysis, it appears that they are included to some extent. In the intact-protein analysis, the measured masses for polypeptides containing pigments are as follows: 17993.9 Da for ApcA, 19822.1 Da for ApcB, 18058.9 Da for CpcA in CPC1, 18058.9 Da for CpcB in CPC1, 21436.7 Da for CpcB in CPC1, and 21436.9 Da for CpcB in CPC2.

To test the molecular assembly, purified proteins of APC, CPC1, and CPC2 were analyzed by size-exclusion column chromatography, as shown in Supplementary Fig. S3.

The main peaks for APC, CPC1, and CPC2 are located at a position slightly laterthan 134 kDa, which corresponds to the trimer molecule (approximately 110 kDa) of the heterodimer composed of the alpha and beta chains. For CPC1 and CPC2, a second peak appears between 443 and 143 kDa, which corresponds to the hexamer molecule of the heterodimer. It appears that the trimer molecule dissociated from the hexamer is observed due to dilution during gel filtration analysis, where the protein concentration at eluted fractions is estimated to be 300 times lower than that for 2D electronic spectroscopy.

To further confirm the molecular assembly, purified proteins of APC, CPC1, and CPC2 were analyzed using negative-staining transmission electron microscopy, as shown in Supplementary Fig. S4. The protein solution was placed on the carbon-coated electron microscopy grid with 0.5 μM in Buffer T, and the placed solution was removed by permeating into torn filter paper. The grid was stained with 2% ammonium molybdate (wt/vol.). Samples were observed under a transmission electron microscope (JEM-1400, JEOL, Japan) at 100 kV at the Life Science Center for Survival Dynamics, University of Tsukuba. The results indicate that APC, CPC1, and CPC2 form a trimer or hexamer, since the dominant particles are approximately 100 nm in diameter. However, we cannot exclude the possibility that APC may include ApcA, ApcB, and ApcC' complexes in addition to dominant ApcA, ApcB, and ApcE' complexes, and CPC1 and CPC2 include trimer molecules even in the conditions of 2D electronic spectroscopy.

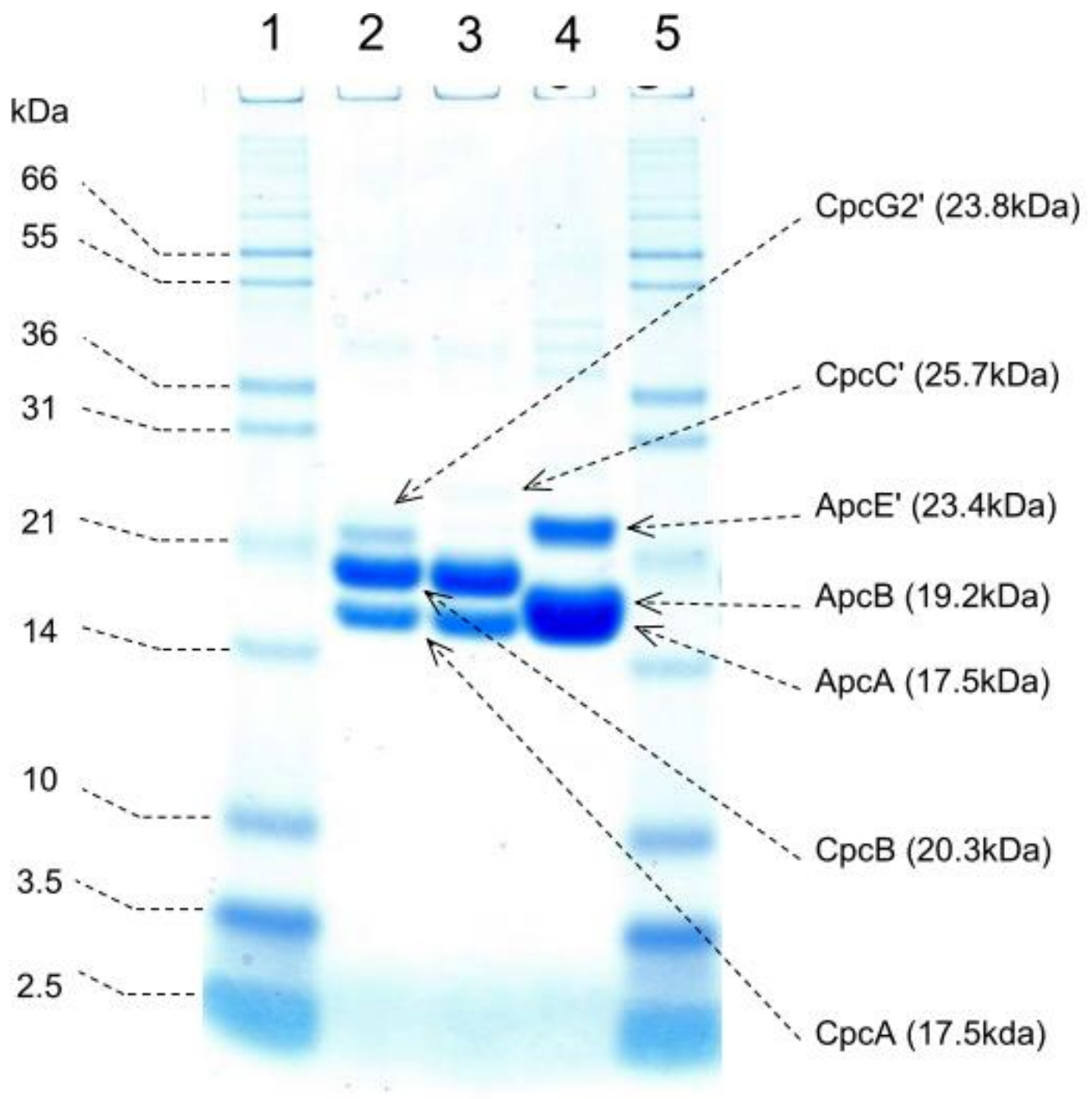

**Supplementary Fig. S2 | SDS-PAGE analysis.** NuPAGE Bis-Tris Precast Gel (Thermo Fisher Scientific) and MES buffer were used for the analysis. Mark12 Unstained Standard was loaded in lanes 1 and 5; CPC2, CPC1, and APC were loaded in lanes 2, 3, and 4, respectively. The gel was stained with EzStain AQua (ATTO) after fixation for 3 hours by 5% glutaraldehyde.

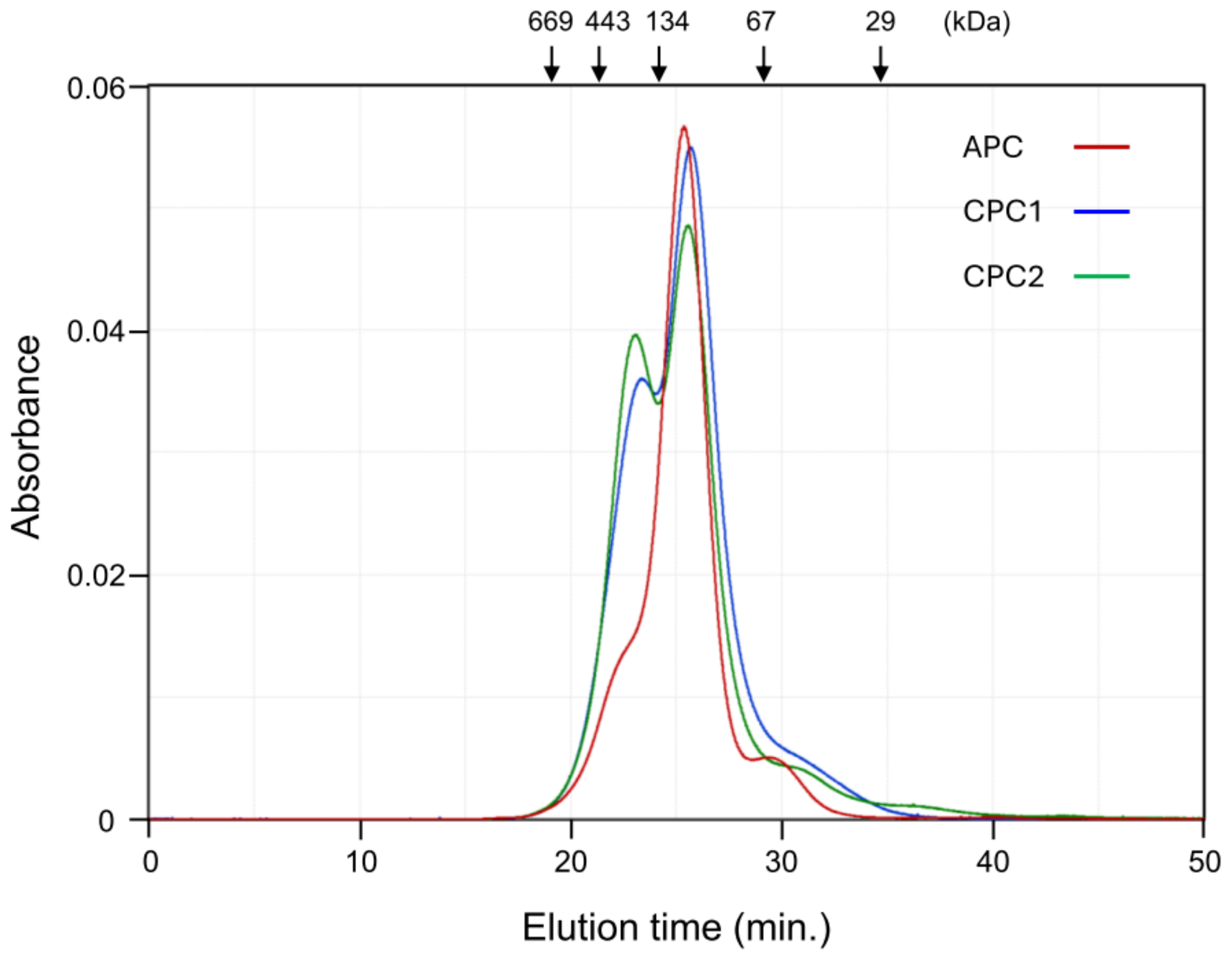

**Supplementary Fig. S3 | Size-exclusion column chromatography.** A Superdex 200 10/300GL gel filtration column was equilibrated with Buffer T. A sample with a volume of 0.03 mL was injected and eluted at 0.5 mL/ml flow rate, and the protein was monitored by measuring the absorbance at 280 nm. Arrows indicate the elution times of bovine thyroglobulin (669 kDa, 18.97 min), apoferritin (443 kDa, 21.41 min), dimer of bovine serum albumin (134 kDa, 24.15 min), bovine serum albumin (67 kDa, 27.5 min), and carbonic anhydrase (29 kDa, 34.23 min).

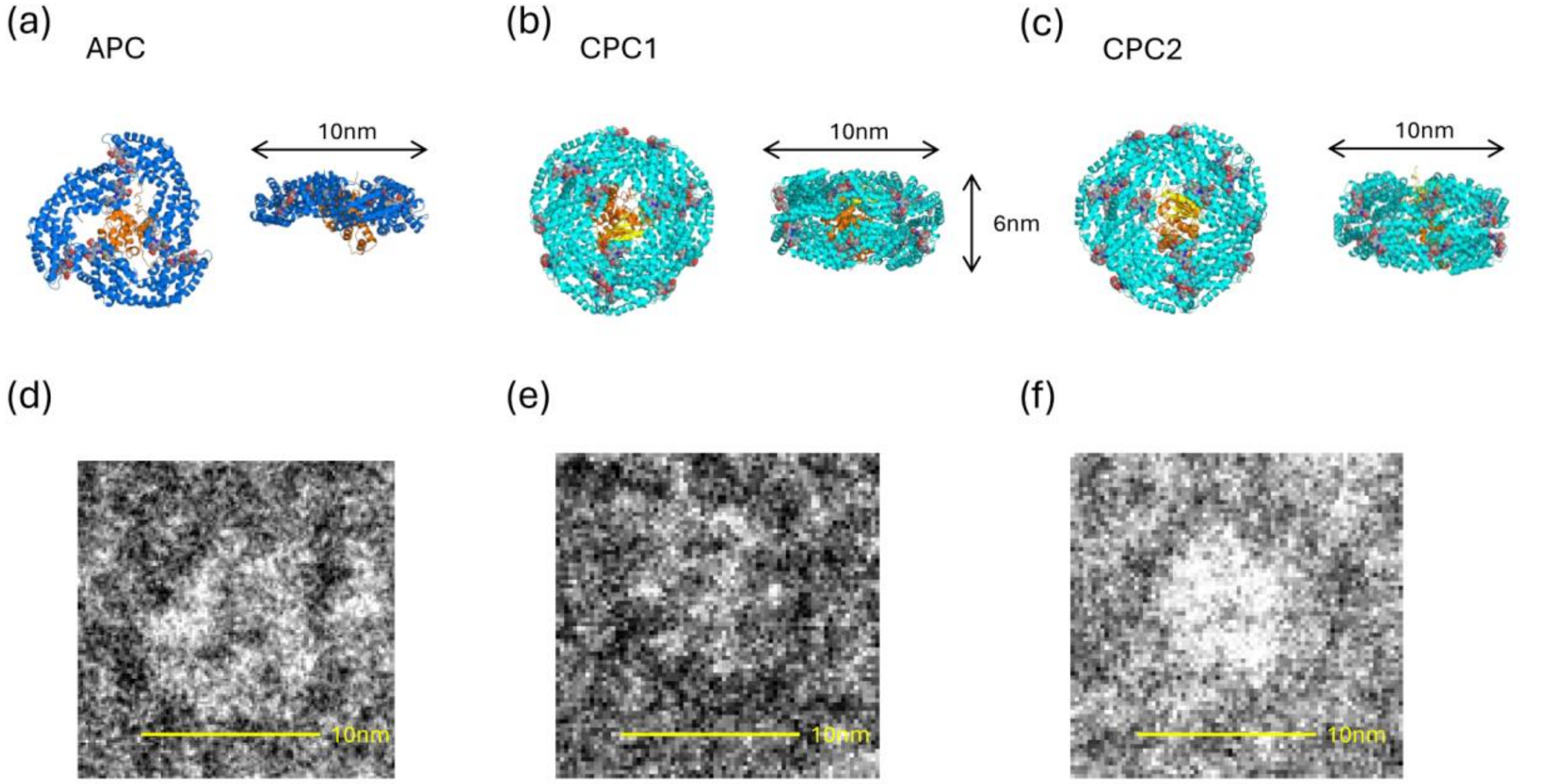

**Supplementary Fig. S4 | Negative-staining transmission electron microscopy.** (a), (b), and (c) show the structures of the supposed protein molecules, APC (ApcA, ApcB, and ApcE' complexes), CPC1 (CpcA, CpcB, CpcD', and CpcC' complexes), and CPC2 (CpcA, CpcB, CpcG1', and CpcC'[2] complexes) represented by ribbon models. (d), (e), and (f) show photos taken by electron microscopy for APC, CPC1, and CPC2, respectively.

## 3. Experimental apparatus for 2D electronic spectroscopy

We produced sub-10-femtosecond visible pulses using a Yb:KGW laser (Light Conversion, Pharos-SP-10W-200kHz, 1030 nm, 1.0 mJ, 10 kHz, 190 fs) followed by two pulse compression stages, as reported previously.[8] The full width at half maximum of the measured pulse duration was 7 fs. The pulse energy after the pulse compression section was approximately 4.7 μJ. The long-term power fluctuation was measured to be 0.26% (standard deviation) for 25 hours.

The 2D electronic spectroscopy (2D-ES) measurement system is a version of 2D infrared spectroscopy[9-11] extended to the visible spectral region in order to investigate the electronic state of molecules. The basic concepts of 2D-ES are summarized in the references.[12, 13] The experimental setup employed in this study is shown in Supplementary Fig. S5(a). First, the sub-10-femtosecond laser pulse with slight negative chirp was split into two pulses by the beam splitter (BS). One of the pulses passed through the computer-controlled translational stage to scan the pump-probe delay time, $T$. The other pulse was used to generate the double pump pulses. The time interval between the double pump pulses, $\tau$, in 2D-ES measurement was scanned with sub-femtosecond temporal resolution by the Translating-Wedge-Based Identical Pulses eNcoding System (TWINS).[14, 15] The first α–BBO crystal with a thickness of 2 mm generated double pulses with polarizations perpendicular to each other. The subsequent pairs of wedge plates of the α–BBO crystal with a wedge angle of 7° varied the interval of the double pulses. The first wire grid polarizers (WG1) in both the pump and probe paths were used for power adjustment, and the other wire grids (WG2) were used to generate pulses with 45° polarization with respect to the optical table. The chirp was compensated by the pair of chirp mirrors (Layertec, 111346) and the pair of fused-silica wedge plates for pump and probe pulses, respectively. To determine the zero of the coherent time, $\tau$, the interferogram between the two pump pulses partially reflected by WG2 placed just after TWINS was measured by the photodetector (PD) and analyzed by the phasing procedure.[16]

The three pulses were focused on the sample by a concave mirror (CM1) with a focal length of 250 mm. After the sample, the pump pulses were spatially filtered, while the signal and probe lights, which were employed as a local oscillator, were collimated by a concave mirror (CM2). After attenuation by the ND filter, the signal and probe light were introduced to a grating spectrometer built in house for heterodyned detection. The signal and probe pulses were spatially dispersed via a holographic grating with a groove density of 300 grooves/mm and focused on a line-scan CMOS camera (UNiiQA+, e2v) with 4096 pixels.

To achieve a background-free heterodyne detection, a chopping system was employed. The chopper (Ch) was synchronized to the laser pulse and operated at a frequency of 2500 Hz. The spectra were averaged for two laser shots, and the recorded data was assigned to Bins I or II according to the chopper timing, as shown in Supplementary Fig. S5(b). To subtract the background spectrum, the spectrum of S = I − II was calculated. This operation was repeated 120 times at each delay time of $\tau$ and $T$, and the 120 spectra of S were averaged. The time interval of the double pump pulses, $\tau$, was scanned from −60 to 60 fs in the steps of 0.4 fs, and the pump-probe delay time, $T$, was scanned from 0 to 2000 fs in the steps of 4 fs. In total, 480 spectra were measured for each delay time $\tau$ and $T$, and, therefore, a complete set of $7.2 \times 10^7$ (= 301 × 501 × 480) spectra were obtained over the entire course of measurements. Including the waiting time for stage movement, the measurements were completed within three hours.

We characterized the pulse duration of the visible pulses at the sample position using a second-harmonic-generation frequency-resolved optical gating (SHG-FROG). The full width at half maximum of the measured pulse duration was 13 fs, which was significantly long compared to the pulse before the 2D-ES apparatus. This was due to the interference effect in the excessive number of reflections in the chirp mirrors. The beam diameter was measured at 0.18 mm by a knife edge method at the sample position.

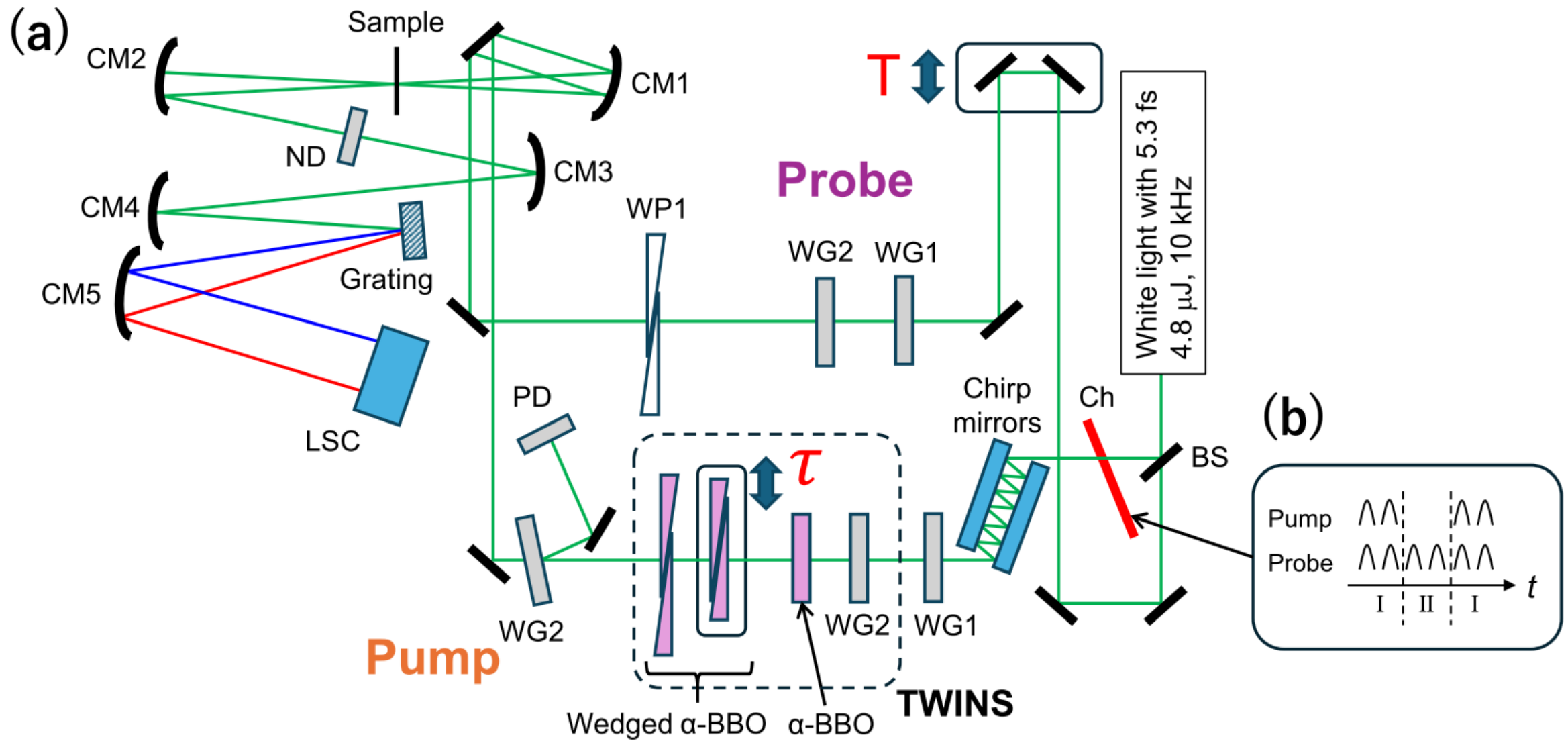

**Supplementary Fig. S5 | Optical layout for 2D-ES.** (a) Schematic diagram of the 2D-ES apparatus. BS: 1 mm thick fused-silica 1:1 beam splitter. Ch: optical chopper. WG1: wire grid polarizers for adjusting the pulse energy. WG2: wire grid polarizers for generating pulses with 45° polarization with respect to the optical table. α–BBO: α–BBO crystal with a thickness of 2 mm. Wedged α–BBO: wedge plate of α–BBO crystal with a wedge angle of 7°. PD: fast photodiode. WP1: a pair of fused-silica dispersion wedge plates with a wedge angle of 2.8°. CM 1, 2, 3, 4, and 5: concave mirrors with focal lengths of 250, 250, 150, 150, and 250 mm, respectively. ND: neutral density filter. Sample: sample cell. Grating: holographic grating with a groove density of 300 grooves/mm. LSC: high-speed line-scan camera with 4096 pixels. Other thick black lines represent silver-coated mirrors. (b) Operation of the chopping system. The figure describes the pulse sequences irradiating the sample.

## 4. Excitation pulse energy dependence

To confirm the absence of non-linear response by a single laser pulse, a series of 2D electronic spectra were measured as a function of the excitation pulse energy. The results shown in Supplementary Fig. S6 are almost independent of the laser pulse energy densities below 60 μJ/cm$^2$. Because the non-linear response was not found from this result, we employed the pulse energy density of 40 μJ/cm$^2$ through this study.

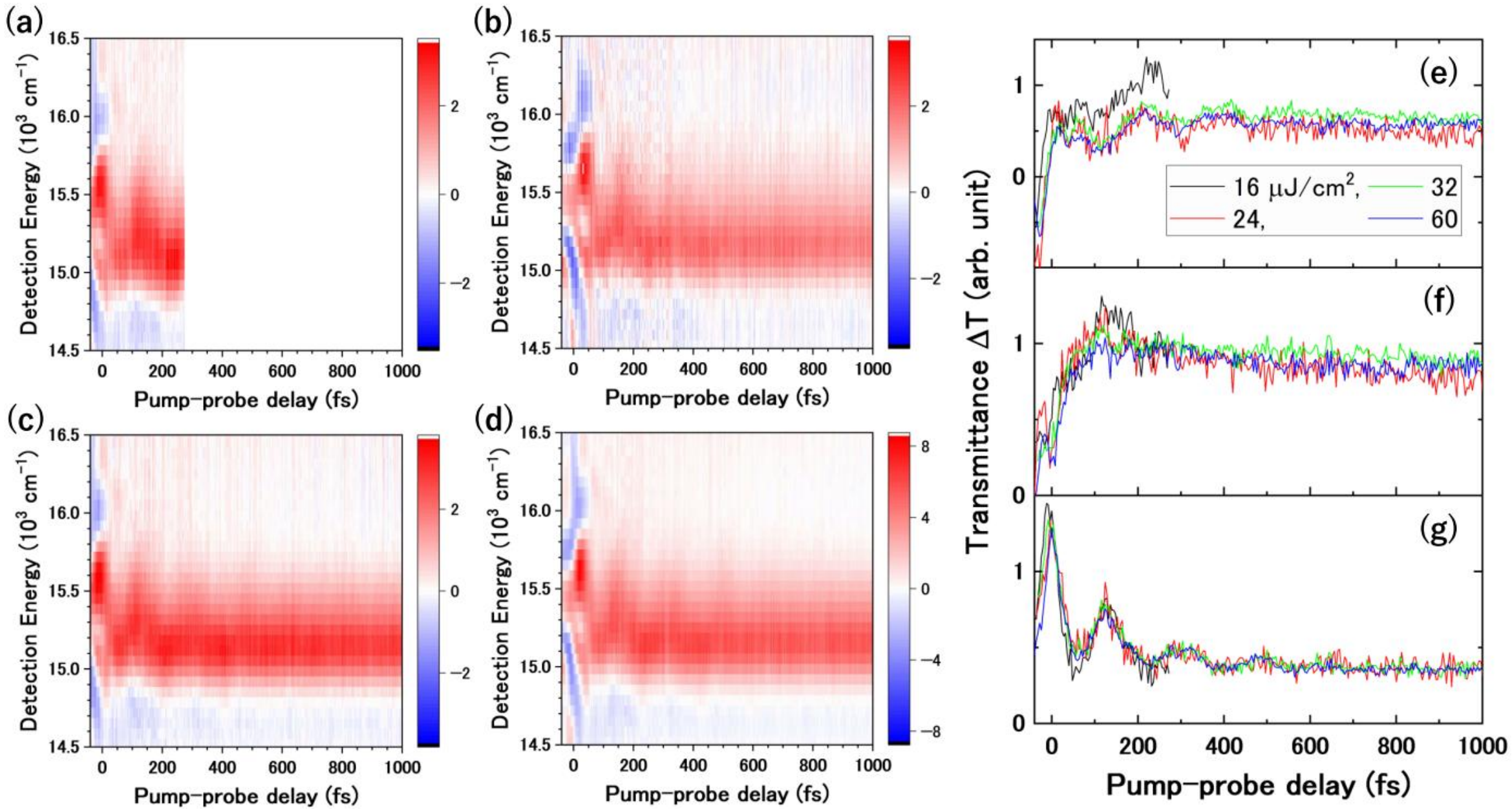

**Supplementary Fig. S6 | Dependence on excitation pulse energy.** Excitation energy selected time-resolved detection signal spectra $S(E_{\mathrm{det}}, T\colon E_{\mathrm{exc}})$ of APC at $E_{\mathrm{exc}}$ = 15650 cm$^{-1}$ with pulse energy densities of (a) 16, (b) 24, (c) 32, and (d) 60 μJ/cm$^2$. The time profiles of the signal amplitude after excitation at detection energies of $E_{\mathrm{det}}$ = 15000, 15250, and 15500 cm$^{-1}$ are shown in (e), (f), and (g), respectively. The black, red, green, and blue lines correspond to the time profiles measured with pulse energy densities of 16, 24, 32, and 60 μJ/cm$^2$, respectively. The intensities are adjusted to make it easier to compare the shapes of the time profiles.

## 5. 2D electronic spectra of CPC2

A series of 2D electronic spectra of CPC2 is shown in Supplementary Fig. S7. The 2D spectra of CPC2 appear very similar to those of CPC1 except for the red shift by 100 $cm^{-1}$ with respect to the axis of the detection energy.

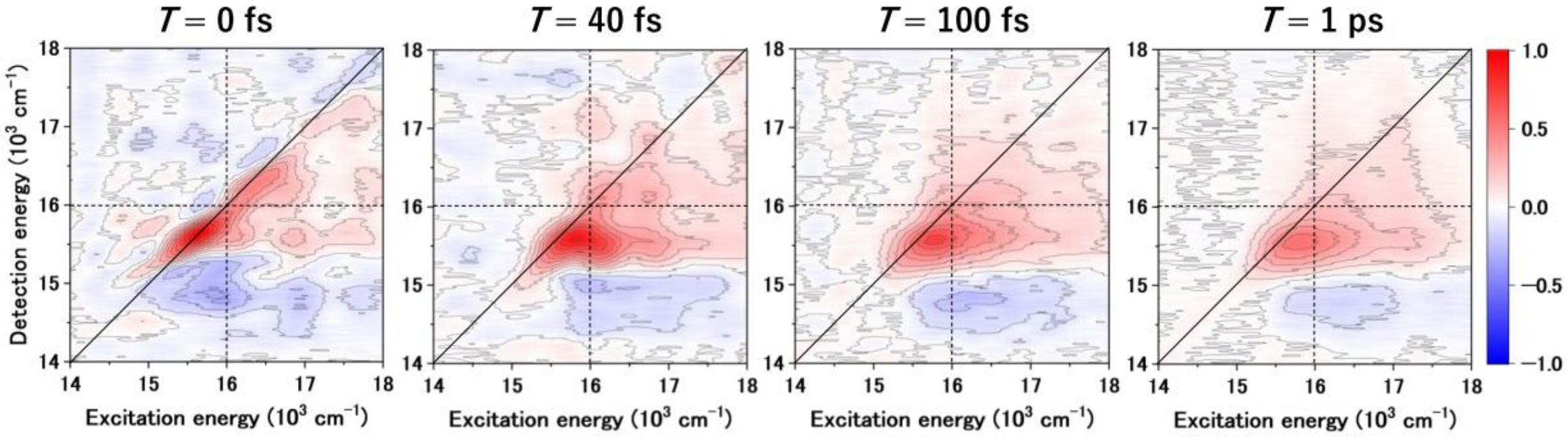

**Supplementary Fig. S7 | Time-resolved 2D electronic spectra of CPC2.** A series of absorptive real-valued 2D spectra of CPC2 measured at pump-probe delay times of $T = 0$ fs, 40 fs, 100 fs, and 1 ps are shown. The dashed lines show the peak energy (16000 $cm^{-1}$) of the absorption spectrum shown in Fig. 1(f) in the main text.

## 6. Excitation-energy-dependent dynamic Stokes shift

Supplementary Fig. S8 compares the excitation-energy-dependent time-resolved detection signal spectra $S(E_{\mathrm{det}}, T: E_{\mathrm{exc}})$ of APC with those of CPC1 and CPC2. No matter what energy APC was excited with, it relaxed to the lowest energy level ($E_{\mathrm{det}}$ = 15150 cm$^{-1}$) within several hundred femtoseconds. On the other hand, in both CPC proteins, the high energy component remains at the long delay time after excitation. In addition to the monotonous decay, the oscillatory component can be clearly found in APC at $E_{\mathrm{exc}}$ = 15500 cm$^{-1}$, which is the peak energy in the absorption spectrum.

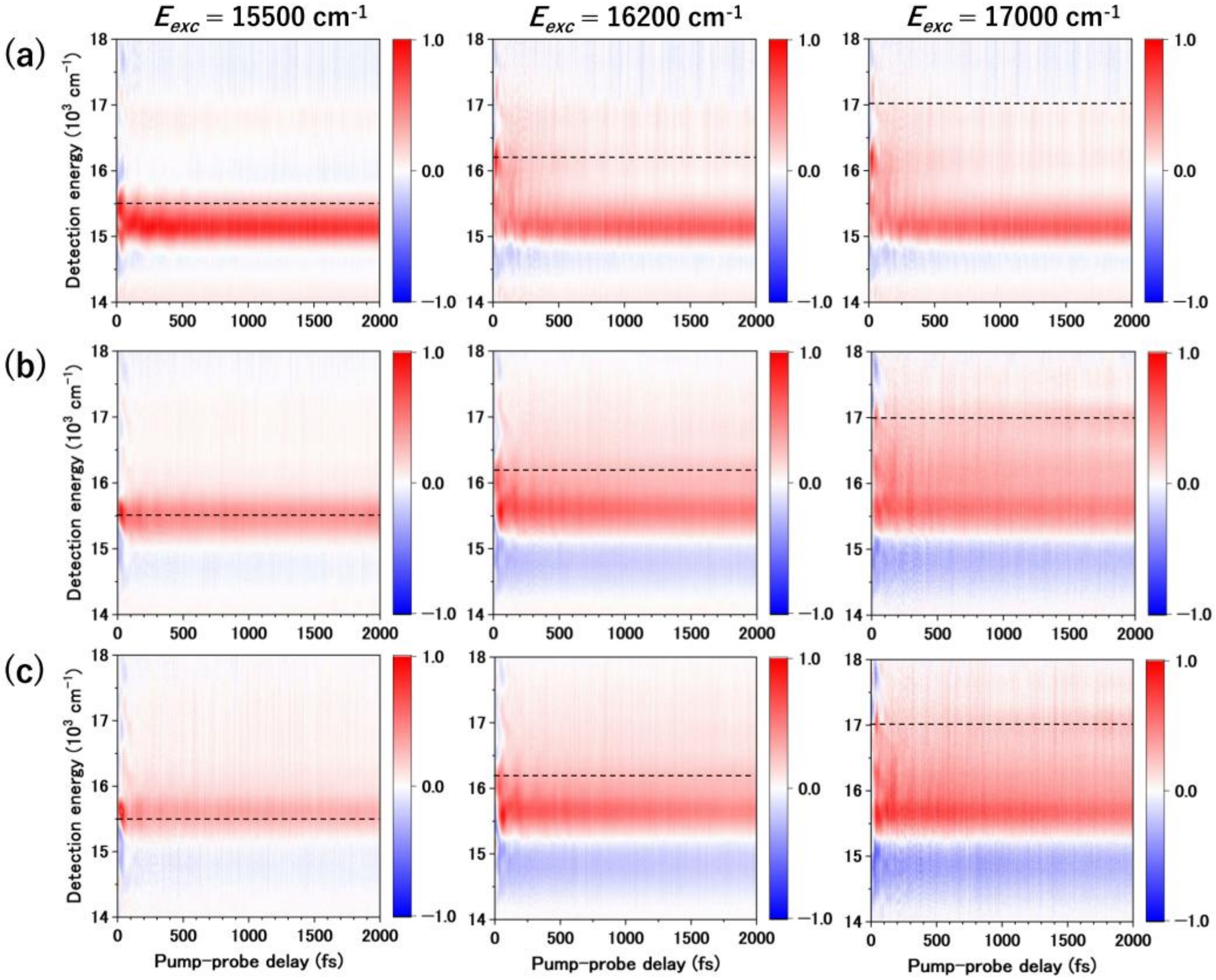

**Supplementary Fig. S8 | Excitation-energy-dependent dynamic Stokes shift.** The excitation-energy-dependent time-resolved detection signal spectra $S(E_{\mathrm{det}}, T: E_{\mathrm{exc}})$ of (a)

APC, (b) CPC2, and (c) CPC1 at $E_{exc}$ = 15500, 16200, and 17000 $cm^{-1}$. The dashed lines indicate the excitation energies.

## 7. Global analysis of 2D electronic spectra

For a quantitative evaluation of energy relaxation and quantum coherence, global analysis is performed for the excitation-energy-dependent time-resolved detection signal spectra $S(E_{\mathrm{det}}, T\colon E_{\mathrm{exc}})$. First, we apply the singular value decomposition to describe $S(E_{\mathrm{det}}, T\colon E_{\mathrm{exc}})$ as a few principal components. However, $S(E_{\mathrm{det}}, T\colon E_{\mathrm{exc}})$ includes the contributions of energy relaxation, the quantum coherence, and the coherent spike within several hundreds of femtoseconds after excitation, and therefore, the decomposition with only a few principal components does not work well at around zero-time delay. In the present work, we employ only the first three principal components by excluding data within a few tens of femtoseconds after excitation from the analysis.

Supplementary Fig. S9 shows the first three principal components of the time profiles $S_{i=1,2,3}(T\colon E_{\mathrm{exc}})$ and the spectra $S_i(E_{\mathrm{det}}\colon E_{\mathrm{exc}})$ obtained by the singular value decomposition of $S(E_{\mathrm{det}}, T\colon E_{\mathrm{exc}} = 15500\ \mathrm{cm}^{-1})$ of APC shown in Fig. 4(a) in the main text. The index $i$ indicates the index of the principal components. The first component is almost stationary within 2 ps. On the other hand, the second and third components show the exponential decay superimposed with the quantum beat. Three principal time profiles are evaluated using the global fitting with a model function:

$$S_i(T\colon E_{\mathrm{exc}}) = A_{\mathrm{f},i}(E_{\mathrm{exc}}) \exp\{-T/\tau_{\mathrm{f}}(E_{\mathrm{exc}})\} + A_{\mathrm{s},i}(E_{\mathrm{exc}}) \exp\{-T/\tau_{\mathrm{s}}(E_{\mathrm{exc}})\} + A_{\mathrm{c},i}(E_{\mathrm{exc}}), \qquad (1)$$

where $\tau_{\mathrm{f}}(E_{\mathrm{exc}})$ and $\tau_{\mathrm{s}}(E_{\mathrm{exc}})$ are the time constants obtained by global fitting at the excitation energy $E_{\mathrm{exc}}$. The pre-factors $A_{\mathrm{f},i}(E_{\mathrm{exc}})$, $A_{\mathrm{s},i}(E_{\mathrm{exc}})$, and $A_{\mathrm{c},i}(E_{\mathrm{exc}})$ are associated with "fast", "slow", and "constant" kinetic compartments as explained below. The evolution-associated difference spectra (EADS) of each compartment are obtained by averaging the principal spectra $S_i(E_{\mathrm{det}}\colon E_{\mathrm{exc}})$ with the pre-factors, for example,

$$S_{\mathrm{f}}(E_{\mathrm{det}}\colon E_{\mathrm{exc}}) = A_{\mathrm{f},1}(E_{\mathrm{exc}}) S_1(E_{\mathrm{det}}\colon E_{\mathrm{exc}}) + A_{\mathrm{f},2}(E_{\mathrm{exc}}) S_2(E_{\mathrm{det}}\colon E_{\mathrm{exc}}) + A_{\mathrm{f},3}(E_{\mathrm{exc}}) S_3(E_{\mathrm{det}}\colon E_{\mathrm{exc}}). \qquad (2)$$

The result of the global fitting for $S(E_{\mathrm{det}}, T\colon E_{\mathrm{exc}} = 15500\ \mathrm{cm}^{-1})$ of APC is shown in Supplementary Fig. S9 as the dashed lines. The decay time constants in the fast and slow compartments are found through global fitting to be $\tau_{\mathrm{f}}$ = 27 fs and $\tau_{\mathrm{s}}$ = 176 fs, respectively.

The last compartment is associated with the stationary state within a time window of 2 ps. In the first compartment, we cannot distinguish between the ultrafast exponential behavior ($\tau_f$ = 27 fs) and the oscillatory component, and therefore, we do not discuss the dynamics of this compartment. Supplementary Fig. S10(a) shows the EADSs of APC in excitation to the lower exciton state at $E_{exc}$ = 15500 cm$^{-1}$. The EADS in the fast compartment shows strong interference between the ultrafast kinetics, quantum beat, and coherent spike as explained above. In the slow compartment, the positive signal *via* the ground state bleaching (GSB) or the stimulated emission (SE) appears at $E_{det}$ = 15370 cm$^{-1}$, which is slightly smaller than the excitation energy. This component corresponds to the vibronic states directly excited by the pump laser, and it decays with a time constant of 176 fs. Note that a negative peak is also seen in the EADS of the slow compartment at 14800 cm$^{-1}$. This is not the decay of the signal *via* the excited state absorption (ESA) but the rising component of the signal from GSB or SE because the signal amplitude of the 2D electronic spectrum is positive at this detection energy, as shown in Figs. 2(a) and 4(a). The time constant $\tau_{slow}$ = 176 fs of the slow compartment represents vibrational energy relaxation on the lower exciton state.

Supplementary Fig. S10(b) shows the EADSs of APC in excitation to the higher exciton state at $E_{exc}$ = 16650 cm$^{-1}$. The time constants of the fast and slow compartments are $\tau_f$ = 4.4 fs and $\tau_s$ = 317 fs, respectively. The positive peak in the slow compartment spectrum is very broad, with a width larger than 1000 cm$^{-1}$, which is the result of the internal conversion from the higher to lower exciton states with the vibrational energy redistribution. The vibrationally excited states in the slow compartment decay to the stationary state to which the positive peak at $E_{det}$ = 15130 cm$^{-1}$ is assigned. This energy is very close to the peak energy shown in Supplementary Fig. S10(a) in the case of the excitation to the lower exciton state at $E_{exc}$ = 15500 cm$^{-1}$.

Supplementary Figs. S10(c) and (d) show the EADSs of CPC1 excited to the lower ($E_{exc}$ = 16150 cm$^{-1}$) and higher ($E_{exc}$ = 16650 cm$^{-1}$) exciton states, respectively. In both cases, the EADS in the fast compartment (black solid lines) possesses a positive peak near the excitation energy. The vibronic states detected as this positive peak quickly decay to the states detected as the broad band spectra (red solid lines) in the slow compartment containing both the positive and negative peaks. Unlike APC, the EADS of CPC1 in the slow compartment is similar to that of the stationary state. This similarity implies that the population of the vibronic states that spread to the wide energy region do not relax to the bottom of the potential energy surface within a time window of 2 ps.

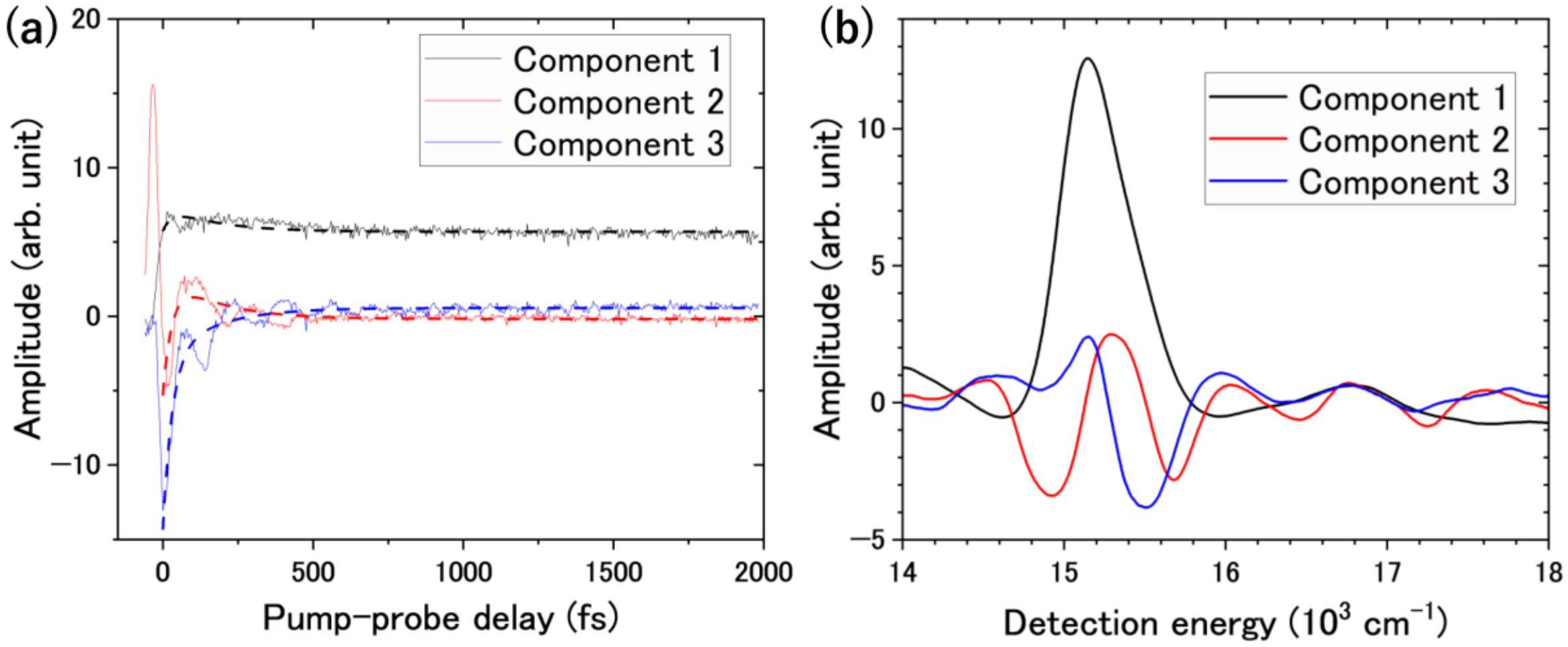

**Supplementary Fig. S9 | Example of singular value decomposition.** (a) First three principal components of the time profiles and (b) spectra obtained by the singular value decomposition of $S(E_{\mathrm{det}}, T\colon E_{\mathrm{exc}} = 15500\ \mathrm{cm}^{-1})$ of APC shown in Fig. 4(a) in the main text. The dashed lines are the results of global fitting to the model function of Eq. (1).

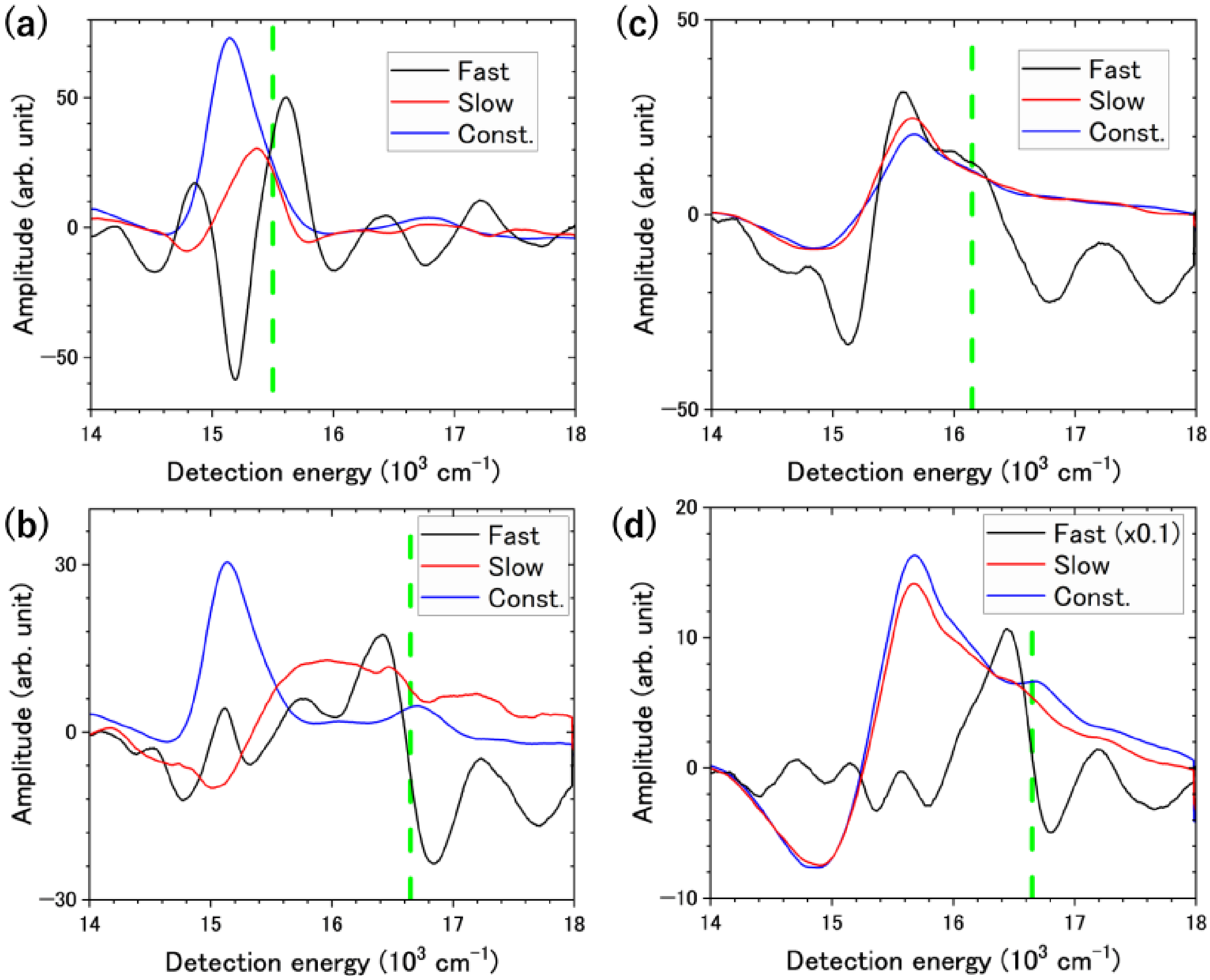

**Supplementary Fig. S10 | Evolution-associated difference spectra.** (a) EADSs of APC excited at 15500 $cm^{-1}$. The black, red, and blue lines are the spectra in the compartments associated with the first, second, and third terms of the model function Eq. (1) used in global fitting. (b) EADSs of APC excited at 16650 $cm^{-1}$. (c) EADSs of CPC1 excited at 16150 $cm^{-1}$. (d) EADSs of CPC1 excited at 16650 $cm^{-1}$. The vertical green dashed lines show the excitation energy.

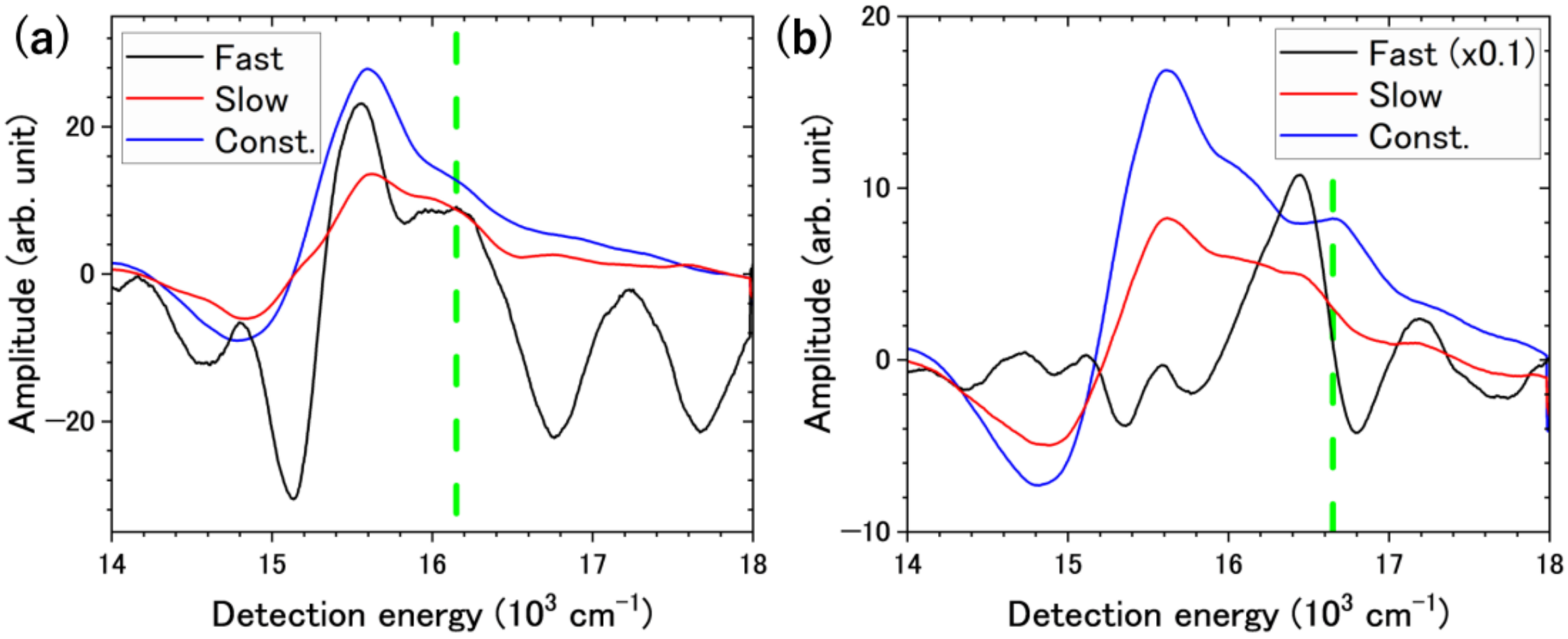

**Supplementary Fig. S11 | EADS of CPC2.** (a) EADS of CPC2 excited at 16150 $cm^{-1}$. The black, red, and blue lines are the compartments associated with the first, second, and third terms of model function Eq. (1) used in global fitting. (b) EADS of CPC2 excited at 16650 $cm^{-1}$. The vertical green dashed lines show the excitation energies.

## 8. Beat-frequency-resolved 2D electronic spectra

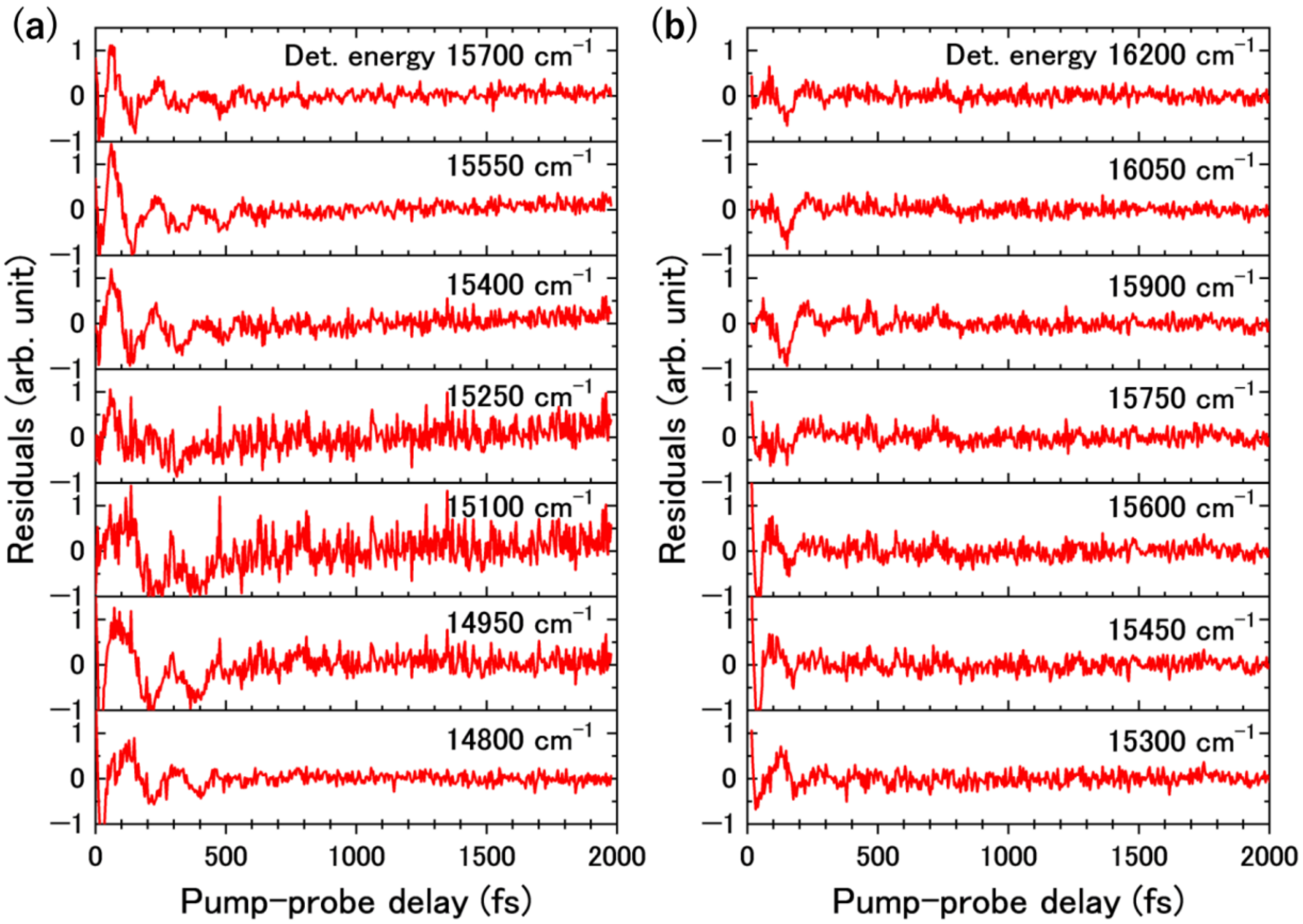

**Supplementary Fig. S12 | Quantum beat signal in the time profiles.** The time profiles of the residuals for (a) APC at the excitation energy of $E_{\mathrm{exc}} = 15500\ \mathrm{cm}^{-1}$ and (b) CPC1 at $E_{\mathrm{exc}} = 16150\ \mathrm{cm}^{-1}$ generated from the nonlinear least-squares fitting explained in Figs. 5(a) and (b), respectively, in the main text. The detection energies are indicated close to the profiles.

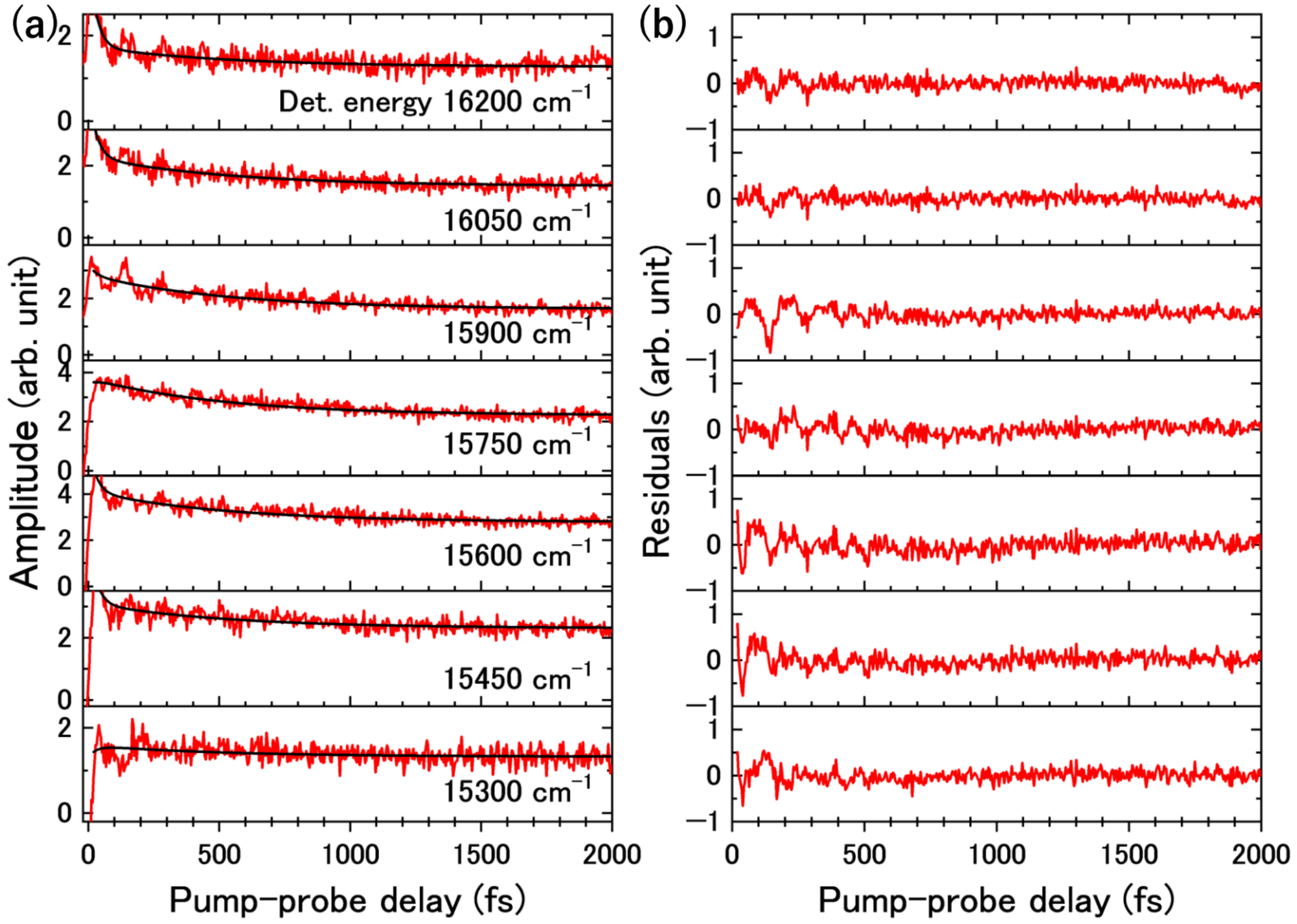

**Supplementary Fig. S13 | Energy relaxation and quantum coherence in the lower exciton state of CPC2.** (a) Time profiles of the signal amplitude after excitation measured for CPC2. The excitation energy is the peak of the absorption spectra, $E_{\mathrm{exc}} = 16150\ \mathrm{cm}^{-1}$. The detection energies are indicated close to the profiles. The red lines show experimental data, and the black lines show the fitting results. (b) Time profiles of the residuals generated from fitting.

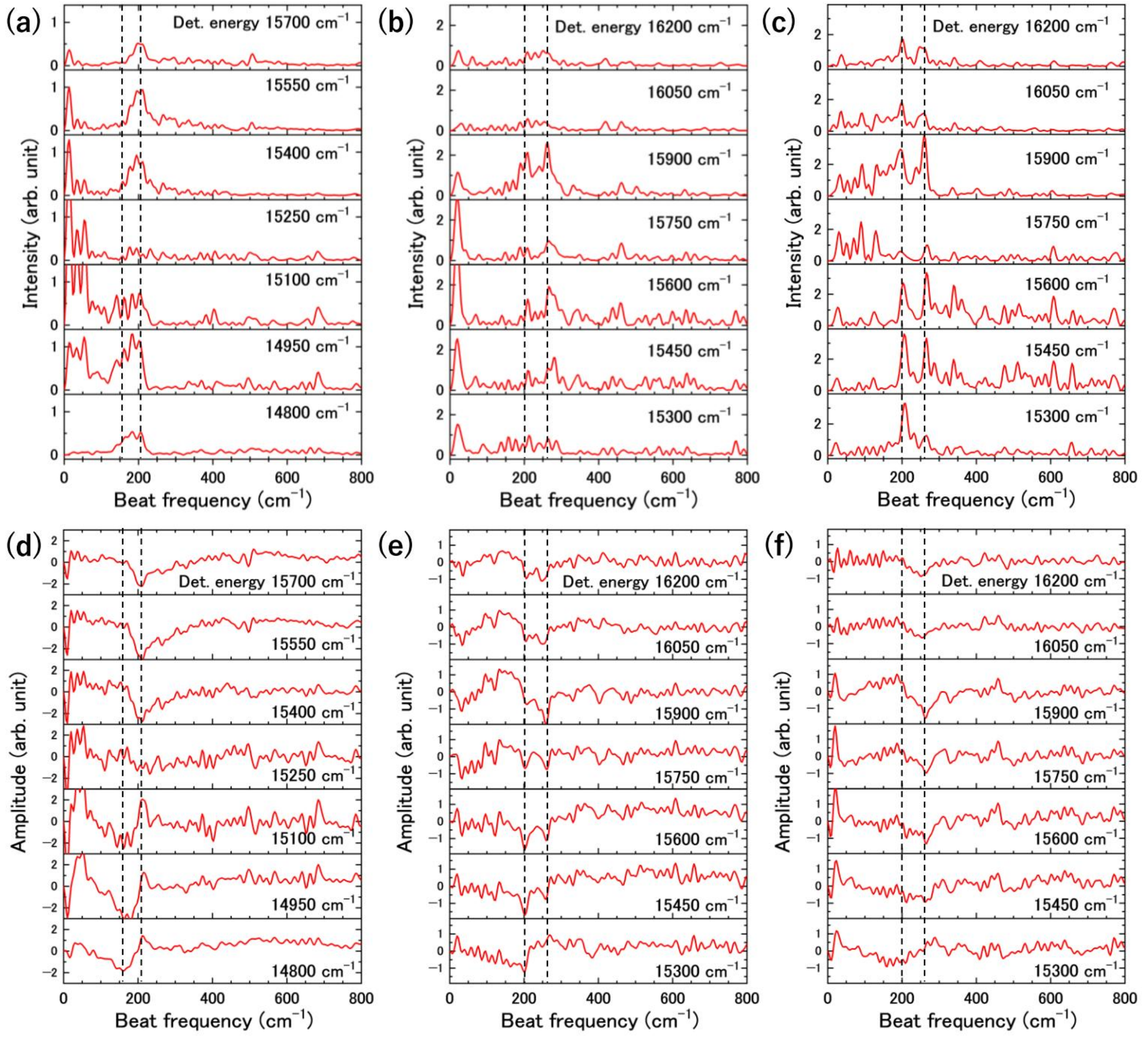

**Supplementary Fig. S14 | Beat-frequency spectra calculated from the Fourier transform of the residual time profiles.** (a), (b), and (c) are the intensity spectra of APC at an excitation energy of $E_{\mathrm{exc}} = 15500\ \mathrm{cm}^{-1}$, of CPC2 at $E_{\mathrm{exc}} = 16150\ \mathrm{cm}^{-1}$, and of CPC1 at $E_{\mathrm{exc}} = 16150\ \mathrm{cm}^{-1}$, respectively. (d), (e), and (f) are the amplitude spectra (the real part of the complex Fourier transform spectra) corresponding to the intensity spectra of (a), (b), and (c), respectively. The detection energies are indicated close to the profiles.

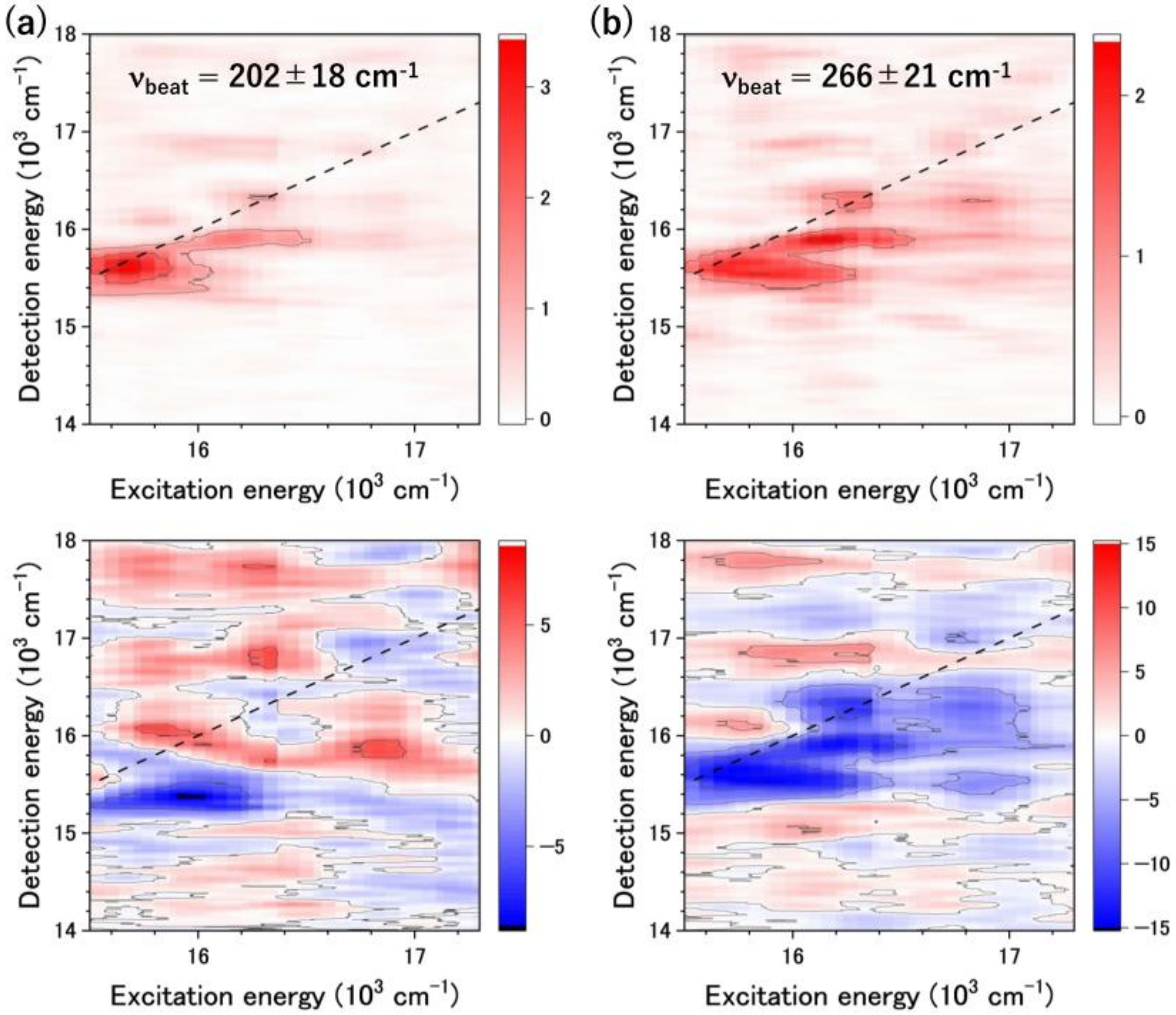

**Supplementary Fig. S15 | Beat-frequency-resolved 2D electronic spectra.** (a) 2D spectra of CPC2 calculated at the beat frequencies $\nu_{\text{beat}}$ = 202 ± 18 cm$^{-1}$. The upper and lower panels are the intensity and amplitude (the real part of the complex values) maps, respectively. The dashed line indicates diagonal. (b) 2D spectra of CPC2 at $\nu_{\text{beat}}$ = 266 ± 21 cm$^{-1}$.

## References

1. K. Kawakami, T. Hamaguchi, Y. Hirose, D. Kosumi, M. Miyata, N. Kamiya and K. Yonekura, "Core and rod structures of a thermophilic cyanobacterial light-harvesting phycobilisome", Nat. Commun. **13**, 3389 (2022).

2. A. McGregor, M. Klartag, L. David and N. Adir, "Allophycocyanin trimer stability and functionality are primarily due to polar enhanced hydrophobicity of the phycocyanobilin binding pocket", J. Mol. Biol. **384**, 406-421 (2008).

3. W. Reuter, G. Wiegand, R. Huber and M. E. Than, "Structural analysis at 2.2 Å of orthorhombic crystals presents the asymmetry of the allophycocyanin–linker complex, AP·$L_C^{7.8,}$ from phycobilisomes of *Mastigocladus laminosus*", Proc. Natl. Acad. Sci. USA **96**, 1363-1368 (1999).

4. J. Nield, P. J. Rizkallah, J. Barber and N. E. Chayen, "The 1.45Å three-dimensional structure of C-phycocyanin from the thermophilic cyanobacterium Synechococcus elongatus", J. Mol. Biol. **141**, 149-155 (2003).

5. L. David, A. Marx and N. Adir, "High-resolution crystal structures of trimeric and rod phycocyanin", J. Mol. Biol. **405**, 201-213 (2011).

6. R. Fromme, A. Ishchenko, M. Metz, S. R. Chowdhury, S. Basu, S. Boutet, P. Fromme, T. A. White, A. Barty, J. C. H. Spence, U. Weierstall, W. Liu and V. Cherezov, "Serial femtosecond crystallography of soluble proteins in lipidic cubic phase", IUCrJ **2**, 545-551 (2015).

7. C. N. Pace, F. Vajdos, L. Fee, G. Grimsley and T. Gray, "How to measure and predict the molar absorption coefficient of a protein", Protein Sci. **4**, 2411-2423 (1995).

8. M. Tsubouchi, N. Ishii, Y. Kagotani, R. Shimizu, T. Fujita, M. Adachi and R. Itakura, "Beat-frequency-resolved two-dimensional electronic spectroscopy: disentangling vibrational coherences in artificial fluorescent proteins with sub-10-fs visible laser pulses", Opt. Express **31**, 6890-6906 (2023).

9. D. B. Strasfeld, S.-H. Shim and M. T. Zanni, "New advances in mid-IR pulse shaping and its application to 2D IR spectroscopy and ground-state coherent control", Adv. Chem. Phys. **141**, 1-28 (2009).

10. S.-H. Shim and M. T. Zanni, "How to turn your pump–probe instrument into a multidimensional spectrometer: 2D IR and Vis spectroscopiesvia pulse shaping", Phys. Chem. Chem. Phys. **11**, 748-761 (2009).

11. P. Hamm and M. Zanni, *Concepts and methods of 2D infrared spectroscopy*. (Cambridge U. Press, Cambridge, 2011).

12. A. Ishizaki and G. R. Fleming, "Quantum coherence in photosynthetic light harvesting", Annu. Rev. Condens. Matter Phys. **3**, 333-361 (2012).

13. D. M. Jonas, "Two-dimensional femtosecond spectroscopy", Annu. Rev. Phys. Chem. **54**, 425-463 (2003).

14. D. Brida, C. Manzoni and G. Cerullo, "Phase-locked pulses for two-dimensional spectroscopy by a birefringent delay line", Opt. Lett. **37**, 3027-3029 (2012).

15. J. Réhault, M. Maiuri, A. Oriana and G. Cerullo, "Two-dimensional electronic spectroscopy with birefringent wedges", Rev. Sci. Instrum. **85**, 123107 (2014).

16. J. Helbing and P. Hamm, "Compact implementation of Fourier transform two-dimensional IR spectroscopy without phase ambiguity", J. Opt. Soc. Am. B **28**, 171-178 (2011).